\documentclass[namedreferences]{solarphysics}

\usepackage[figuresright]{rotating}
\usepackage[hyperref,optionalrh,solaromanenum]{spr-sola-addons}
\usepackage{graphicx}
\usepackage{txfonts}
\usepackage{natbib}
\usepackage{amssymb}
\usepackage{amsbsy}
\usepackage{color}
\usepackage{nicefrac}
\usepackage{booktabs}
\usepackage{threeparttable}

\usepackage{caption}
\DeclareCaptionLabelSeparator{myspace}{\rule{4pt}{0pt}}
\hypersetup{
    final=true,
    pageanchor=true,
    colorlinks=true,
    breaklinks=true,
    linkcolor=blue,
    citecolor=blue,
    urlcolor=blue,
    pdfpagemode=UseNone,
    pdftitle={Wavelength Dependence of Image Quality Metrics and Seeing
        Parameters and their Relation to Adaptive Optics Performance},
    pdfauthor={Kamklah et al.},
    pdfsubject={Solar Physics},
    pdfkeywords={Granulation, Photosphere, Chromosphere, Image Restoration, 
        Adaptive Optics, Instrumentation and Data Management}}

\newcommand\phm{\phantom{$-$}}
\newcommand\phn{\phantom{0}}
\newcommand\degr{\ensuremath{^\circ}}

\definecolor{myBlue}{rgb}{.2,.7,.9}

\definecolor{applegreen}{rgb}{0.55, 0.71, 0.0}

\definecolor{mvGreen}{cmyk}{0.992,0.,0.083, 0.525}

\definecolor{com}{rgb}{0.698039, 0.0941176, 0.133333}

\definecolor{myGreen}{rgb}{.2,.8,.2}

\begin{document}

\begin{article}

%
%

\begin{opening}


\title{Wavelength Dependence of Image Quality Metrics and Seeing Parameters 
    and their Relation to Adaptive Optics Performance}

\author[addressref={aff1,aff2,aff3,aff4}, corref, email={rkamlah@aip.de}]
    {\inits{R.\ }\fnm{R.\ }\lnm{Kamlah}\orcid{0000-0003-2059-585X}}
\author[addressref={aff1}]
    {\inits{M.\ }\fnm{M.\ }\lnm{Verma}\orcid{0000-0003-1054-766X}}    
\author[addressref={aff1,aff5}]
    {\inits{A.\ }\fnm{\\A.\ }\lnm{Diercke}\orcid{0000-0002-9858-0490}}
\author[addressref={aff1}]
    {\inits{C.\ }\fnm{C.\ }\lnm{Denker}\orcid{0000-0002-7729-6415}}

\address[id=aff1]{Leibniz-Institut f{\"u}r Astrophysik Potsdam (AIP), 
    An der Sternwarte 16, 
    14482 Potsdam, Germany}
\address[id=aff2]{Technische Hochschule Wildau,
    Hochschulring 1,
    15745~Wildau, Germany}
\address[id=aff3]{Technische Hochschule Brandenburg,
    Magdeburger Stra{\ss}e 50,
    14770~Brandenburg an der Havel, Germany} 
\address[id=aff4]{Tor Vergata University of Rome,
    School of Mathematical, Physical and Natural Sciences, Via della Ricerca Scientifica 1, 00133 Rome, Italy}
\address[id=aff5]{Universit{\"a}t Potsdam,
    Institut f{\"u}r Physik und
    Astronomie, Karl-Liebknecht-Stra{\ss}e 24/25,
    14476~Potsdam, Germany}

\runningauthor{Kamlah et al.}
\runningtitle{Wavelength Dependence of Image Quality Metrics and Seeing
    Parameters}


\begin{abstract}
Ground-based solar observations are severely affected by Earth's turbulent atmosphere. As a consequence, observed image quality and prevailing seeing conditions are closely related. Partial correction of image degradation is nowadays provided in real-time by adaptive optics (AO) systems. In this study, different metrics of image quality are compared with parameters characterizing the prevailing seeing conditions, \textit{i.e.}, \textit{Median Filter Gradient Similarity} (MFGS), \textit{Median Filter Laplacian Similarity} (MFLS), Helmli-Scherer mean, granular rms-contrast, differential image motion, and Fried-parameter $r_0$. The quiet-Sun observations at disk center were carried out at the \textit{Vacuum Tower Telescope} (VTT), \textit{Observatorio del Teide} (OT), Iza{\~n}a, Tenerife, Spain. In July and August 2016, time-series of short-exposure images were recorded with the \textit{High-resolution Fast Imager} (HiFI) 
at various wavelengths in the visible and near-infrared parts of the spectrum. Correlation analysis yields the wavelength dependence of the image quality metrics and seeing parameters, and \textit{Uniform Manifold Approximation and Projection} (UMAP) is employed to characterize the seeing on a particular observing day. In addition, the image quality metrics and seeing parameters are used to determine the field-dependence of the correction provided by the AO system. Management of high-resolution imaging data from large-aperture, ground-based telescopes demands reliable image quality metrics and meaningful characterization of prevailing seeing conditions and AO performance. The present study offers guidance how to retrieve such information \textit{ex post facto}. 
\end{abstract}


    
\end{opening}

%
%

\section{Introduction}\label{s:intro} 

Operating large-aperture solar telescopes within Earth’s turbulent atmosphere leads, next to instrument inherent limitations (\textit{e.g.}, diffraction limit), to a reduced image quality, which is a result of wavefront deformations induced by variations of the refractive index resulting from temperature fluctuations and wind motions. The degradation of observing quality, referred to as ``seeing'', is an ongoing field of research with major branches of site characterization \citep[\textit{e.g.},][]{Brandt1982, Zirin1988b, Vernin1994, Hill2006, Verdoni2007}, real-time correction with adaptive optics (AO) systems \citep[\textit{e.g.},][]{Rimmele2011}, and image restoration techniques that are applied \textit{ex post facto} \citep[\textit{e.g.},][]{Loefdahl2007, Denker2015, Loefdahl2016}. 

In statistical studies, single parameters are commonly used to characterize the time dependence of seeing at an observatory site. Nevertheless, most seeing parameters can be computed for each isoplanatic patch, for example, the granular rms-contrast \citep[e.g.,][]{vonderLuehe1987} can be derived for regions sufficiently large to minimize the contributions of individual granules. One of the relevant seeing parameters is the Fried-parameter $r_0$ \citep{Fried1966}, which describes the diameter of coherent turbulent cells within the telescope pupil. Differential image motion \citep[DIM,][]{Sarazin1990} is another well established seeing parameter used to describe rubber-sheet wavefront deformations.

The coherence time of daytime seeing is just a few milliseconds. Therefore, \citet{Martin1987} investigated the impact of exposure time on image quality and emphasized that very short exposure times are needed to `freeze' differential and rigid image motion, a prerequisite for all shift-and-add methods \citep{Hunt1983, Loefdahl2010} aimed at preserving spatial resolution while enhancing photometric precision. However, many image restoration techniques derive seeing information from time-series of short-exposure images \citep{vonderLuehe1993, Loefdahl2002, vanNoort2006}. Separating seeing from object information requires that the observed object does not change while capturing the time-series. This poses significant constraints on image restoration, when observing rapidly evolving solar features with larger and larger telescope apertures. Evaluating three-dimensional spatio-temporal data cubes of solar granulation, recorded at disk center in a sufficiently short time so that the scene on the Sun is not changing, makes it possible to assess the influence of seeing on the recorded data including its field dependence after AO correction.

Image quality metrics provide an alternative method to evaluate seeing. In contrast to seeing parameters, which cover the limited scope of imaging through a turbulent medium, such a metric can be applied to any kind of digital image. \citet{Deng2015} introduced a no-reference, blind, and objective image quality metric called \textit{Median Filter Gradient Similarity} (MFGS) to solar physics, which was modified and extended by \citet{Denker2018b}. The latter method implements smoother and directional invariant derivatives by using the Scharr operator \citep{Scharr2007} for computing gradients. \citet{Popowicz2017} investigated several image quality metrics by simulating seeing effects and applying them to high-resolution solar observations from space. The Helmli-Scherer mean \citep[HSM,][]{Helmli2001} was identified as computationally efficient and MFGS performed well under good seeing conditions, that is, $D/r_0 < 4$, where $D$ is the diameter of the telescope aperture. An evaluation and comparison of gradient-based and other image quality metrics was carried out by \citet{Xue2014}, who also investigated their suitability for real-time image evaluation as needed in AO applications. The good performance of gradient-based operators is in agreement with the results provided by \citet{Pertuz2013}, who conclude that Laplacian-based operator shows the best overall performance under good seeing conditions. However, further investigations of these operators are needed, in particular their dependence on the observed texture requires further scrutiny.

In previous studies \citep[\textit{e.g.},][]{Denker2005b, Denker2007a, Berkefeld2010, Deng2015, Popowicz2017}, various methods were suggested and evaluated to characterize seeing, its impact on image quality, and the performance of AO systems. The motivation of the present study is to utilize some of these methods to assess the performance of the \textit{Vacuum Tower Telescope} \citep[VTT,][]{vonderLuehe1998}, its AO system, and large-format, high-cadence sCMOS imagers. Another objective is to investigate the wavelength dependence of image quality metrics and seeing parameters. 

The article is organized as follows. Observations and data processing are presented in Section~\ref{SEC02}. Image quality metrics and estimated seeing parameters are introduced in Section~\ref{SEC03}. Discussion of radial dependence of AO corrections, correlation analysis of wavelength dependence, and correlation between various image quality metrics and seeing parameters are included in Sections~\ref{SEC04}, \ref{SEC05}, and \ref{SEC06}, respectively. A novel way to classify and find relationships between metrics and parameters using an unsupervised machine learning algorithm is put forward in Section~\ref{SEC07}. Discussions and conclusions are presented in Sections~\ref{SEC08} and~\ref{SEC09}, respectively.

%
%

\section{Observations and Data Processing}\label{SEC02}

The observation were carried out in July and August 2016 using the VTT with an aperture of $D = 70$~cm, which is equipped with the \textit{Kiepenheuer Adaptive Optics System} \citep[KAOS,][]{vonderLuehe2003, Berkefeld2010} for real-time image correction. The observing setup featured two synchronized sCMOS imagers, which are usually integrated in the \textit{High-resolution Fast Imager} \citep[HiFI,][]{Denker2018a, Denker2018b} at the 1.5-meter GREGOR solar telescope \citep{Schmidt2012, Denker2012}. The image processing and restoration techniques, which are required to process and evaluate the data, are provided by the ``sTools'' data processing pipeline \citep{Kuckein2017a}. The target of this investigation was solar granulation in quiet-Sun regions near disk center, exploiting its uniform and isotropic characteristics for benchmarking image quality metrics and seeing parameters. On 2016 July~16 \& 17, active regions NOAA~12565 and 12567 were located in the northern hemisphere in proximity to disk center \citep[see \textit{SolarMonitor},][]{Gallagher2002a}. Thus, the disk-center observations were shifted by about 100~arcsec to the south. In consequence, enhanced small-scale magnetic fields may still be present on these two days. On the following day until and including 2016 August~2, all active regions rotated beyond the solar limb and the disk center was void of any major magnetic activity. The observing characteristics are summarized in Table~\ref{TAB01}.

\begin{table}[t]
\fontsize{7}{8} \selectfont
\begin{threeparttable}
\caption{Observing characteristics in July and August 2016.}
\label{TAB01}
\begin{tabular}{cccccccc}
\toprule
Date & Start & End & Sets & CWL & FWHM & $\Delta t$ & $r_0$ \rule{0mm}{8pt}\\
\midrule
2016-07-16 & 13:46:34~UT & 13:48:55~UT & \phn 8 & 532.82~nm & \phn 9.3~nm & \phn 1.2~ms & (6.86$\pm$0.90)~cm \rule{0mm}{7pt}\\
2016-07-17 & 09:22:25~UT & 09:28:24~UT &     16 & 532.82~nm & \phn 9.3~nm & \phn 1.0~ms & (6.81$\pm$0.74)~cm \\
2016-07-19 & 09:58:03~UT & 10:08:35~UT &     32 & 861.01~nm &     10.1~nm & \phn 2.0~ms & (7.52$\pm$0.73)~cm \\
2016-07-24 & 08:20:59~UT & 08:33:32~UT &     21 & 396.80~nm & \phn 0.1~nm &     30.0~ms & (5.51$\pm$1.05)~cm \\
2016-07-25 & 08:57:36~UT & 09:03:20~UT &     18 & 632.71~nm &     10.0~nm & \phn 2.5~ms & (6.07$\pm$0.88)~cm \\
2016-07-27 & 07:53:11~UT & 08:01:02~UT &     24 & 710.88~nm & \phn 9.7~nm & \phn 3.5~ms & (6.46$\pm$0.72)~cm \\
2016-08-01 & 08:04:44~UT & 08:10:58~UT &     20 & 396.80~nm & \phn 1.1~nm & \phn 6.0~ms & (6.17$\pm$0.93)~cm \\
2016-08-02 & 07:52:19~UT & 07:58:37~UT &     20 & 590.64~nm & \phn 9.7~nm & \phn 2.0~ms & (5.81$\pm$0.77)~cm \\
\bottomrule
\end{tabular}
\footnotesize
\begin{tablenotes}
\item The central wavelength (CWL) and full width at half maximum (FWHM) refer to the interference filters used with Camera~2. The corresponding values for the inference filter in front of Camera~1 are $\lambda$430.7~nm (Fraunhofer G-band) and FWHM = 1.12~nm, respectively. The exposure time $\Delta t$ was exactly the same for both cameras. Eight time-series with a total of 159 datasets were analyzed, which contain $2 \times 100$ short-exposure images each. Mean values and standard deviations of the Fried-parameter $r_0$ are computed for the daily observations in the Fraunhofer G-band.
\end{tablenotes}
\end{threeparttable}
\normalsize
\end{table}

The optical setup in inner tower of the VTT just behind the exit of the AO system consisted of (1) a transfer optics providing an image scale of 0.06~arcsec pixel$^{-1}$ or 9.1~arcsec~mm$^{-1}$, which is about twice the image scale at the primary focus, (2) a 20/80 beamsplitter, and (3) interference filters right in front of the sCMOS cameras. Both cameras were triggered by a programmable timing unit (PTU) with nanosecond precision, ensuring that the exposure times and intervals were exactly the same for both cameras. Neutral density filters were used to adjust the light levels such that both cameras had a count rate corresponding to between two-thirds and three-quarters of the full-well capacity. A detailed description of the synchronized sCMOS imaging system is given in \citet{Denker2018a, Denker2018b}. Camera~1 was always equipped with an interference filter for observing the Fraunhofer G-Band $\lambda$430.7~nm, so that the observations with Camera~2 were always tied to this reference wavelength. This makes it possible to compare the image quality metrics and seeing parameters across the visible and near-infrared wavelength ranges. Mostly broad-band interference filters were used with Camera~2, which provide proxies of the continuum radiation at the given wavelength. Much narrower interference filters were used for observations of the inner part and core of the strong chromospheric absorption line Ca\,\textsc{ii}\,H $\lambda$396.8~nm. The sCMOS cameras operate in the wavelength range 370\,--\,1100~nm, with a maximum quantum efficiency of 57\% at $\lambda$600~nm.

The focus of this investigation is on disk-center observations. However, center-to-limb observations were taken as well but their analysis is deferred to a forthcoming study. The sCMOS detectors have $2560 \times 2160$ pixels, which yield a field of view (FOV) of about 153.6~arcsec $\times$ 129.6~arcsec. The observations comprise several time-series with a small number of datasets. Initially, each observed dataset consists of 500 short-exposure images, which were acquired at a rate of 49~Hz. The cadence was about 20~s, which includes some overhead for saving the data. Thus, the longest time-series of 32 datasets lasted around 10~min. In preparation for archiving HiFI data, basic dark and flat-field corrections are carried out already on-site, and only the best 100 out of 500 images are selected and saved for further analysis. The frame selection is based on image quality, which is determined with the MFGS metric after 2$\times$2-pixel binning to save computation time. 

Initializing the coelostat mirrors before observations and daily switching of interference filters of Camera~2 lead to a small misalignment of the beam of light on both cameras. In addition, imperfections of optical elements and interference filters as well as the absence of rotation stages for the cameras introduce minor differences in image scale and a minor rotation of the image plane. The misalignment errors of camera were determined from the average images of each dataset by a correlation analysis. Image shifts, rotational offsets, and differences in image scale are given in Table~\ref{TAB02}. 

\begin{table}[t]
\footnotesize
\begin{threeparttable}
\caption{Misalignment parameters for Camera~2.}
\label{TAB02}
\begin{tabular}{ccccc}
\toprule
Date & $\Delta x$ & $\Delta y$ & $\Delta \theta$ & $\Delta s$ \rule{0mm}{10pt}\\
\midrule
2016-07-16 &   $-7$~pixel & \phn\phm 1~pixel & 0.377\degr & $-0.26$\%\rule{0mm}{10pt}\\
2016-07-17 & \phm 5~pixel &  \phn $-9$~pixel & 0.374\degr & $-0.28$\%\\
2016-07-19 & \phm 3~pixel &      $-29$~pixel & 0.373\degr & \phm 0.73\%\\
2016-07-24 &   $-3$~pixel & \phn\phm 1~pixel & 0.380\degr & $-1.11$\%\\
2016-07-25 & \phm 4~pixel &      $-27$~pixel & 0.366\degr & $-0.37$\%\\
2016-07-27 &     12~pixel &      $-41$~pixel & 0.384\degr & \phm 0.25\%\\
2016-08-01 & \phm 0~pixel & \phn\phm 5~pixel & 0.389\degr & \phm 0.31\%\\
2016-08-02 &     13~pixel &      $-29$~pixel & 0.386\degr & $-0.14$\%\\
\bottomrule
\end{tabular}
\begin{tablenotes}
\item Camera~2 is shifted by $\Delta x$ and $\Delta y$ pixels and rotated by $\Delta \theta$. Its image scale differed by $\Delta s$ with respect to the nominal image scale of $s = 0.06$~arcsec~pixel$^{-1}$. 
\end{tablenotes}
\end{threeparttable}
\normalsize
\end{table}

Significant values of image shift $\Delta x$ and $\Delta y$ are most likely caused by the 45\degr-turning mirror, which directs the vertical light beam from the telescope to the horizontal optical benches. The large changes are introduced by the turning mirror when compensating small offsets resulting from adjustment of the coelostat mirrors. In contrast, the changes in image scale $\Delta s$ are related to optical manufacturing defects, which affect the surface quality of the interference filters. Because of these imperfections the filters act as weak lenses leading to small  changes in magnification. To take into account all these minor flaws of the optical setup, a careful alignment of the images from both cameras must be carried out in data processing, since the goal is comparing the field dependence of image quality metrics and seeing parameters. The morphological differences were too large between photospheric G-band and chromospheric narrow-band Ca\,\textsc{ii}\,H images. Thus, the alignment was based on images of a resolution target placed  in the science focus just after the AO system. Changes during the observing day can be safely neglected given the short time-series. The linear correlation between pairs of averaged images approaches 90\%. Finally, only the images of Camera~2 were scaled, rotated, and shifted. Taking the alignment errors into account, all images were cropped by 40~pixels on each side, resulting in 2480$\times$2080-pixel images with a FOV of 148.8~arcsec $\times$ 124.8~arcsec.

\begin{figure}[t]
\centering
\includegraphics[width=\textwidth]{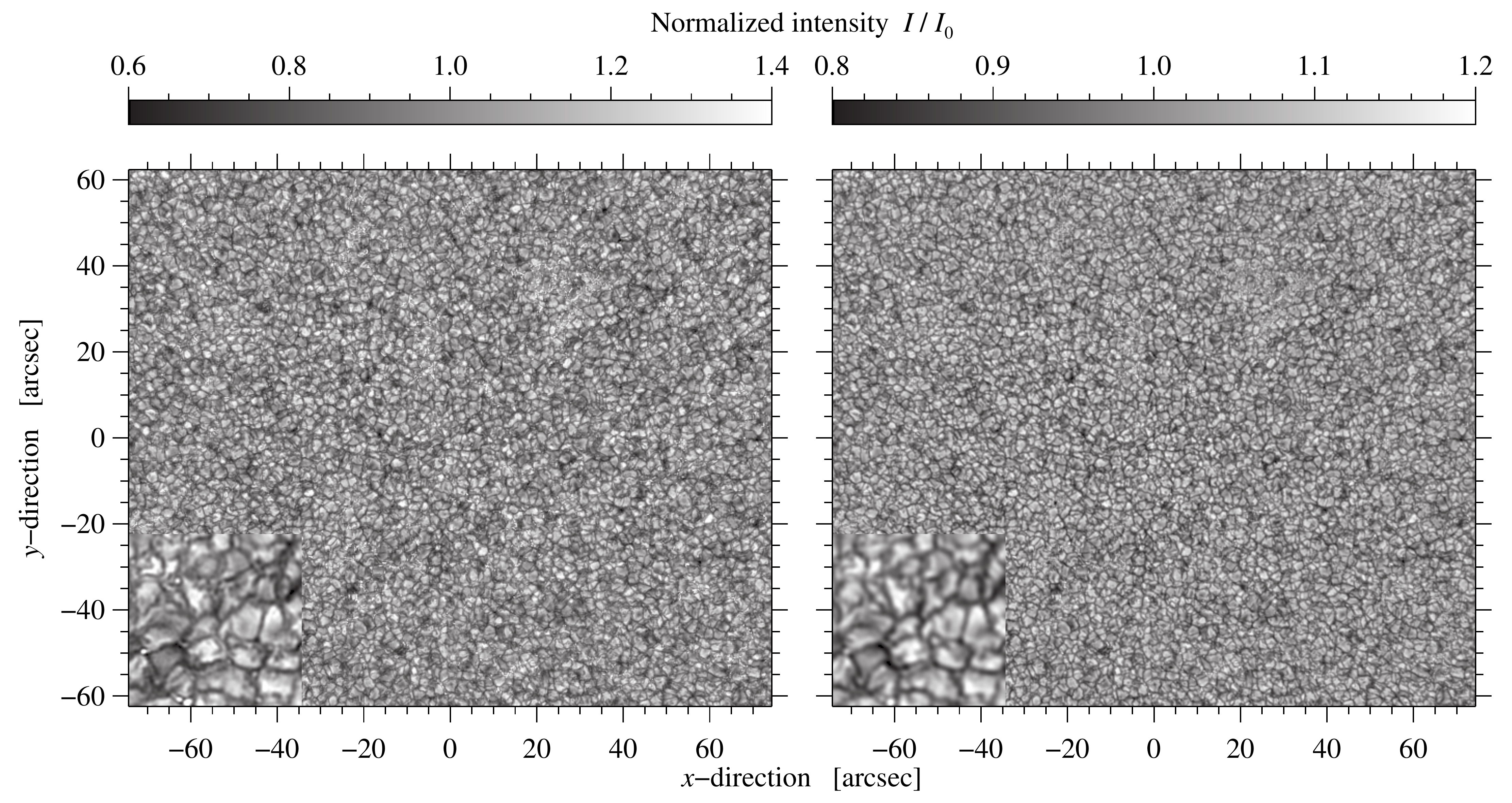}
\caption{Speckle-restored G-band $\lambda$430.7~nm image (\textit{left}) and
    near-infrared $\lambda$861.0~nm continuum image (\textit{right}) observed at 10:01:47~UT on 2016 July~19. The small insert in the lower left corner of each panel displays a 10~arcsec $\times$ 10~arcsec region taken from the center of the FOV at a 4$\times$ magnification.}
\label{FIG01}
\end{figure}

As an example, Figure~\ref{FIG01} shows speckle restored data recorded on 2016 July~19 \citep[see][for details of the image restoration]{Denker2018b}. This dataset serves as reference for the results presented in Section~\ref{SEC03}. The left panel displays the quiet-Sun in the Fraunhofer G-band recorded with Camera~1 and the right panel an image captured by Camera~2 using an interference filter with a central wavelength $\lambda861.0~$nm. A zoom-in of the granulation pattern with a size of 10~arcsec $\times$ 10~arcsec is included in the lower left corner of each panel to demonstrate that in the obtained data the photospheric fine structures are resolved, reaching the diffraction limit of the telescope.

After careful alignment of both imaging channels and considering that the images are taken strictly simultaneously, the derived image quality metrics and seeing parameters can be directly compared. The continuous G-band observations provide the tie that binds the multi-wavelengths observations together. The image quality metrics and seeing parameters are computed for overlapping subfields with $80 \times 80$ pixels, which corresponds to 4.8~arcsec $\times$ 4.8~arcsec or 3.5~Mm $\times$ 3.5~Mm on the solar surface. This small FOV corresponds roughly to the size of the isoplanatic patch in daytime seeing conditions at good sites for solar observations. The seeing parameters were determined for an equidistant grid of $121 \times 101$ grid points with a spacing of 20~pixels or 1.2~arcsec. Conscious oversampling will assist later in computing the radial averages of the image quality metrics and seeing parameters.

%
%

\section{Image Quality Metrics and Seeing Parameters}\label{SEC03}

The main difference between image quality metrics and seeing parameters is that the former require only an image whereas the latter involve image sequences. Thus, assessing the seeing conditions and the field dependence of the AO correction at the same time demands image sequences, \textit{i.e.}, three-dimensional spatio-temporal data cubes. All metrics and parameters of a dataset are computed for the same $121 \times 101$ stacks of subfields with $80 \times 80 \times 100$ pixels so that they can be directly compared.

Image quality metrics are typically defined across two-dimensional intensity distributions. However, their application to three-dimensional stacks of isoplanatic patches is straightforward. The smaller number of pixels contained in an isoplanatic patch is partly compensated by the number of images in a dataset. The seeing parameters, \textit{i.e.}, the Fried-parameter $r_0$ and DIM, can only be derived for the three-dimensional stacks of isoplanatic patches. The computation of image displacements and of power spectra is based on Fourier transforms. Noisy data leads for some isoplanatic patches to erroneous results, which are replaced by a 3$\times$3-pixel median filter applied to the $121 \times 101$ grid points.


\subsection{Granular Contrast}

Using the properties that solar granulation is uniform and isotropic, at least in proximity to disk center, the granular rms-contrast is a simple way to quantify the influence of seeing on image quality \citep{Ricort1981, vonderLuehe1987}. The center-to-limb variation and wavelength dependence of the granular rms-contrast is given by \citet{WedemeyerBoehm2009b}, among others, who compare three-dimensional radiative (magneto-)hydrodynamic simulations with observations of the \textit{Solar Optical Telescope} \citep[SOT,][]{Tsuneta2008} onboard the Japanese \textit{Hinode} mission \citep{Kosugi2007}. Accurate wavelength-dependent point-spread functions (PSFs) of \textit{Hinode}/SOT were derived by \citet{WedemeyerBoehm2008} from solar eclipse observations and a Mercury transit. After PSF and scattered light correction, the granular rms-contrast rises significantly \citep[\textit{e.g.},][]{Danilovic2008, Mathew2009}, which has, for example, profound implications for inversions of Stokes profiles \citep[see][]{vanNoort2013}. However, accurate photometry is not essential for this investigation, which will only add another layer of complexity. 

In Figure~\ref{FIG02}, the top panels represent false-color images of the granular rms-contrast for the G-band and near-infrared $\lambda$861.0~nm filters (see Figure~\ref{FIG01}). The G-band rms-values for Camera~1 are in the range 4\,--\,8\%, whereas the range of 2\,--\,4\% is a factor two lower for the near-infrared rms-values. Large rms-contrasts are found near the AO lock point  close to the center of the FOV. The drop of rms-contrasts from the center to the periphery of the FOV is more pronounced in the near-infrared compared to the G-band. The rms-contrasts at the periphery are a good proxy for observations without AO systems. Visual inspection reveals that the rms-contrasts are considerably higher in a region of approximately 40~arcsec around the AO lock point, which is much larger than the isoplanatic patch size. In both panels, bright granules outshine the local neighborhood. They leave an imprint in adjacent isoplanatic patches because of their significant overlap.


\subsection{Helmli-Scherer Mean}

The Helmli-Scherer mean (HSM) compares the intensity at each pixel $I(x,y)$ with the mean intensity $\mu(x,y)$ in a certain neighborhood \citep{Helmli2001} -- in the present case a 3$\times$3-pixel neighborhood, though larger neighborhoods can also be considered depending on the application. The HSM is defined as
\begin{equation}
R(x,y) = \left\{    
\begin{array}{ll}
\displaystyle\frac{\mu(x,y)}{I(x,y)}, \quad & \mu(x,y) \geq I(x,y)\smallskip\\
\displaystyle\frac{I(x,y)}{\mu(x,y)} & \mathrm{otherwise},
\end{array}
\right.
\label{EQN01}
\end{equation}
where $R(x,y) \equiv 1$ in regions with constant intensities and $R(x,y) > 1$ in regions containing contrast features. The mean $\mu(x,y)$ is implemented straightforward as a smoothing operation with a boxcar average, which is applied to all images in a sequence. Individual $R(x,y)$ values are not meaningful so that the HSM must be computed over a certain two-dimensional region or in the present case, for each stack of subfields. The larger the $R(x,y)$ values, the larger is the local contrast. The good HSM performance was emphasized by \citet{Pertuz2013} who compiled a comprehensive list of gradient-, Laplacian-, wavelet-, statistics-, and discrete-cosine-transform-based operators, among others. These operators are commonly used in shape-from-focus applications \citep{Nayar1994} and can also serve as image quality metrics.

\begin{figure}[t]
\centering
\includegraphics[width=\textwidth]{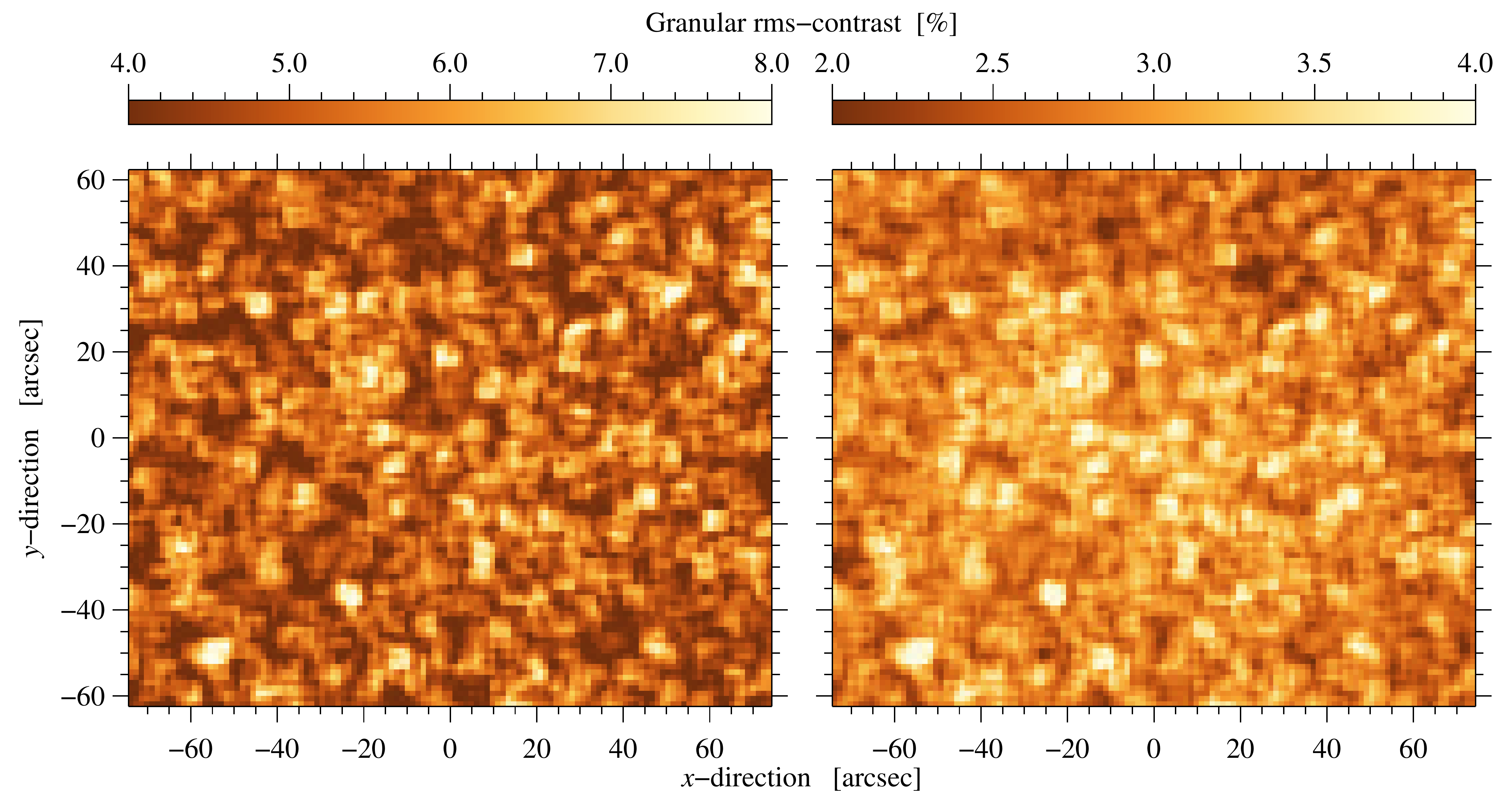}
\includegraphics[width=\textwidth]{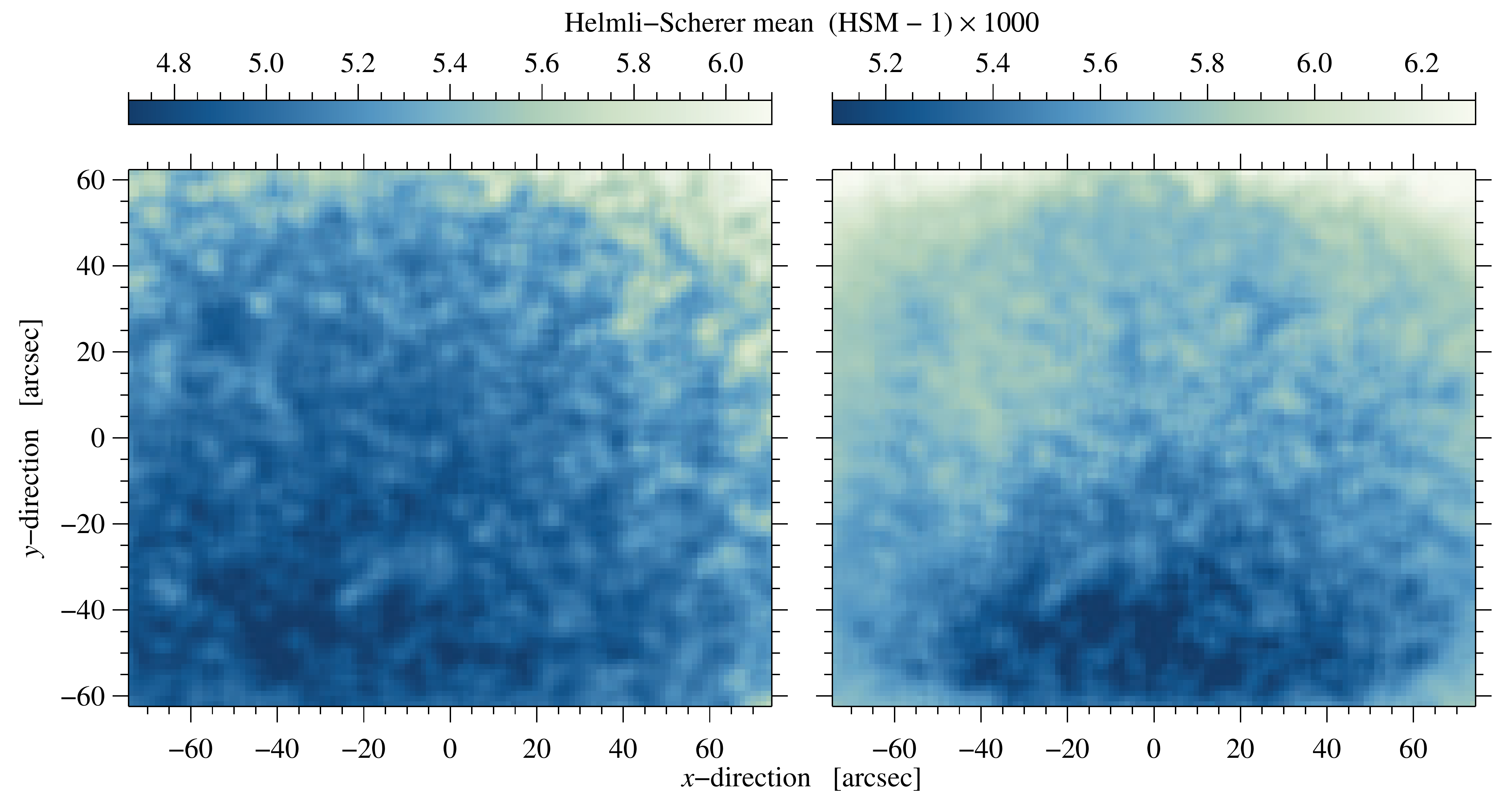}
\caption{Two-dimensional maps of granular rms-contrast (\textit{top}) and
    Helmli-Scherer mean (\textit{bottom}) corresponding to Figure~\ref{FIG01}. The HSM values were scaled to clearly show the small deviations from unity.}
\label{FIG02}
\end{figure}

The field dependence of the HSM is displayed in the lower panels of Figure~\ref{FIG02}. The left panel shows the two-dimensional map for the Fraunhofer G-band and the right panel contains the near-infrared $\lambda861.0$~nm map. The values of the HSM exceed unity only by a few percent, which had to be taken into account for a proper annotation of the scale bars. While the granulation pattern also leaves an imprint in HSM maps, they differ significantly from those of the rms-contrast. A clear radial dependence is not apparent. In both HSM maps high values occur in the corners of the FOV. This trend is more pronounced in the $\lambda861.0$~nm map. A radial trend become only visible when averaging many HSM maps from different observing days. Note that scaling the two-dimensional maps for better visualization is subjective. However, care was taken that the interpretation of the maps is independent of the scaling and the chosen color table.


\subsection{Median Filter Gradient and Laplacian Similarity}

The MFGS image quality metric was introduced by \citet{Deng2015} to evaluate the quality of solar images. It is less dependent on the observed solar scene and compares spatial gradients of raw and median-filtered images, which are denoted by $I_r$ and $I_m$, respectively. The median filter uses a 3$\times$3-pixel neighborhood. The MFGS values are bracketed between zero and unity, thus facilitating in principle a quantitative comparison of images from different instruments and in different wavelength regions, once all dependencies are identified and proper benchmarks are established. Deviating from the original implementation of the MFGS algorithm, Scharr operators \citep{Scharr2007} 
\begin{equation}
G_\mathrm{x} = \left[ \begin{array}{rrr}
-3 & 0 & 3 \\                                       
-10 & 0 & 10 \\
-3 & 0 & 3 \\                                        
\end{array}\right]
\quad \mathrm{and} \quad
G_\mathrm{y} = \left[ \begin{array}{rrr}
3 & 10 & 3 \\                                        
0 & 0 & 0 \\
-3 & -10 & -3 \\                                    
\end{array}\right].
\label{eq:MFGS_y}
\end{equation}
are used in the present study to compute the gradient magnitude \citep[see][]{Denker2018b} 
\begin{equation}
G = \sqrt{(G_x^2 + G_y^2)}.
\end{equation}
Once the gradient magnitudes $G(I_r)$ and $G(I_m)$ are computed for the raw and median-filtered images, respectively, their similarity is computed according to
\begin{equation}
\mathrm{MFGS} = \frac{2\,\sum G(I_r)\,\sum G(I_m)}{(\sum G(I_r))^2 + (\sum G(I_m))^2},
\label{eq:MFGS}
\end{equation}
where the summation typically applies to an entire two-dimensional gradient image. However, in the present case, the summation is carried out for each stack of subfields. 
 
Instead of using the gradient magnitude operator, evaluating the local curvature is another viable option. Thus, a new image quality metric, \textit{i.e.}, the \textit{Median Filter Laplacian Similarity} (MFLS) is introduced in this study, which employs convolution with a Laplacian kernel
\begin{equation}
L = \left[
\begin{array}{ccc}
1 & \phantom{-}1 & \phantom{-}1\\
1 & -8 & \phantom{-}1\\
1 & \phantom{-}1 & \phantom{-}1
\end{array}
\right].
\end{equation}
In comparison to MFGS, MFLS requires fewer mathematical operations because of the point symmetry of $L$, \textit{i.e.}, only one convolution of $L$ with the raw and median filtered images and taking the absolute value. Again, this image quality metric is based on the similarity measure
\begin{equation}
\mathrm{MFLS} = \frac{2\,\sum |L(I_r)|\,\sum |L(I_m)|}{(\sum |L(I_r)|)^2 + (\sum |L(I_m)|)^2}.
\label{eq:MFLS}
\end{equation}

\begin{figure}[t]
\centering
\includegraphics[width=\textwidth]{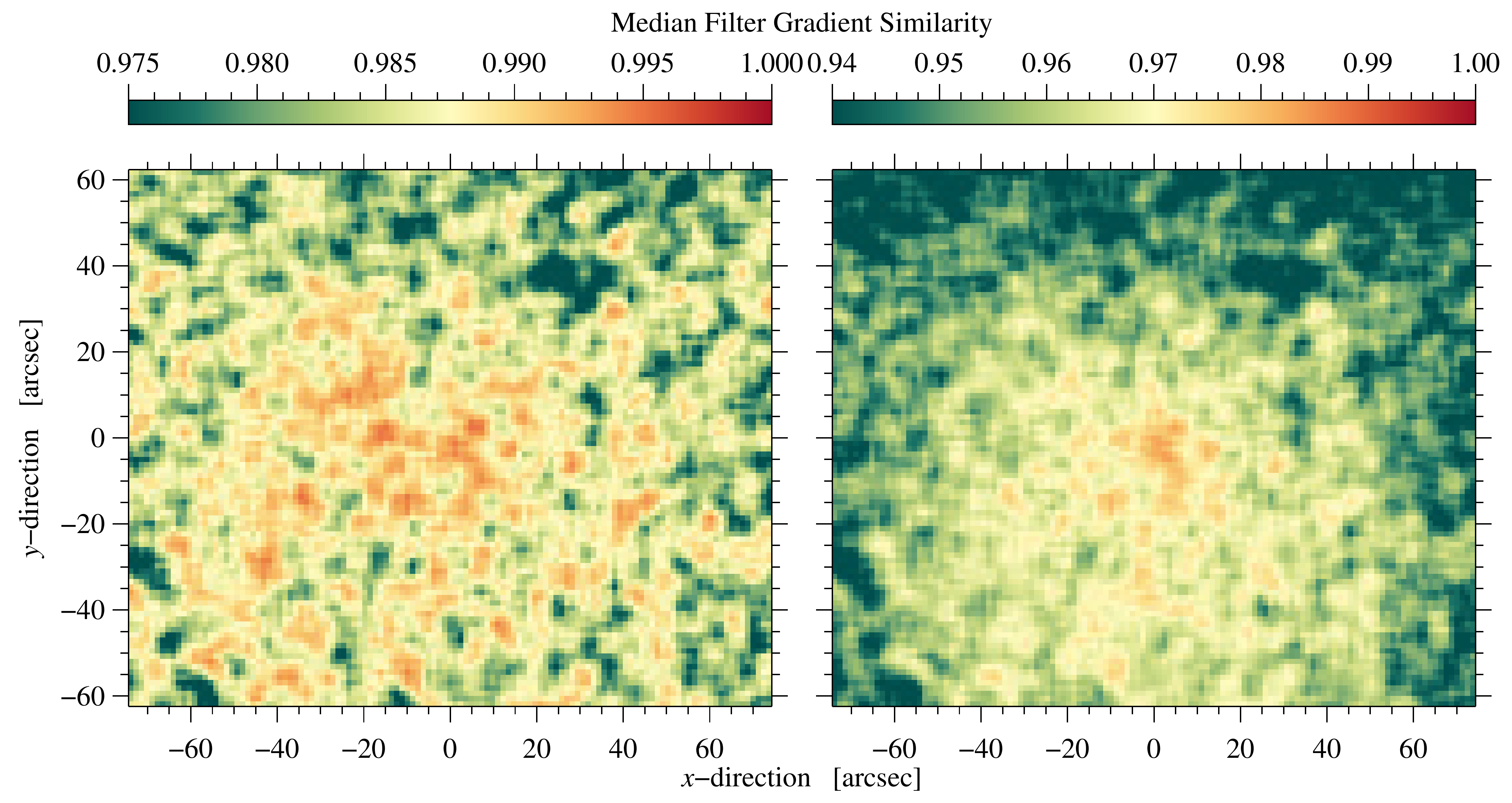}
\includegraphics[width=\textwidth]{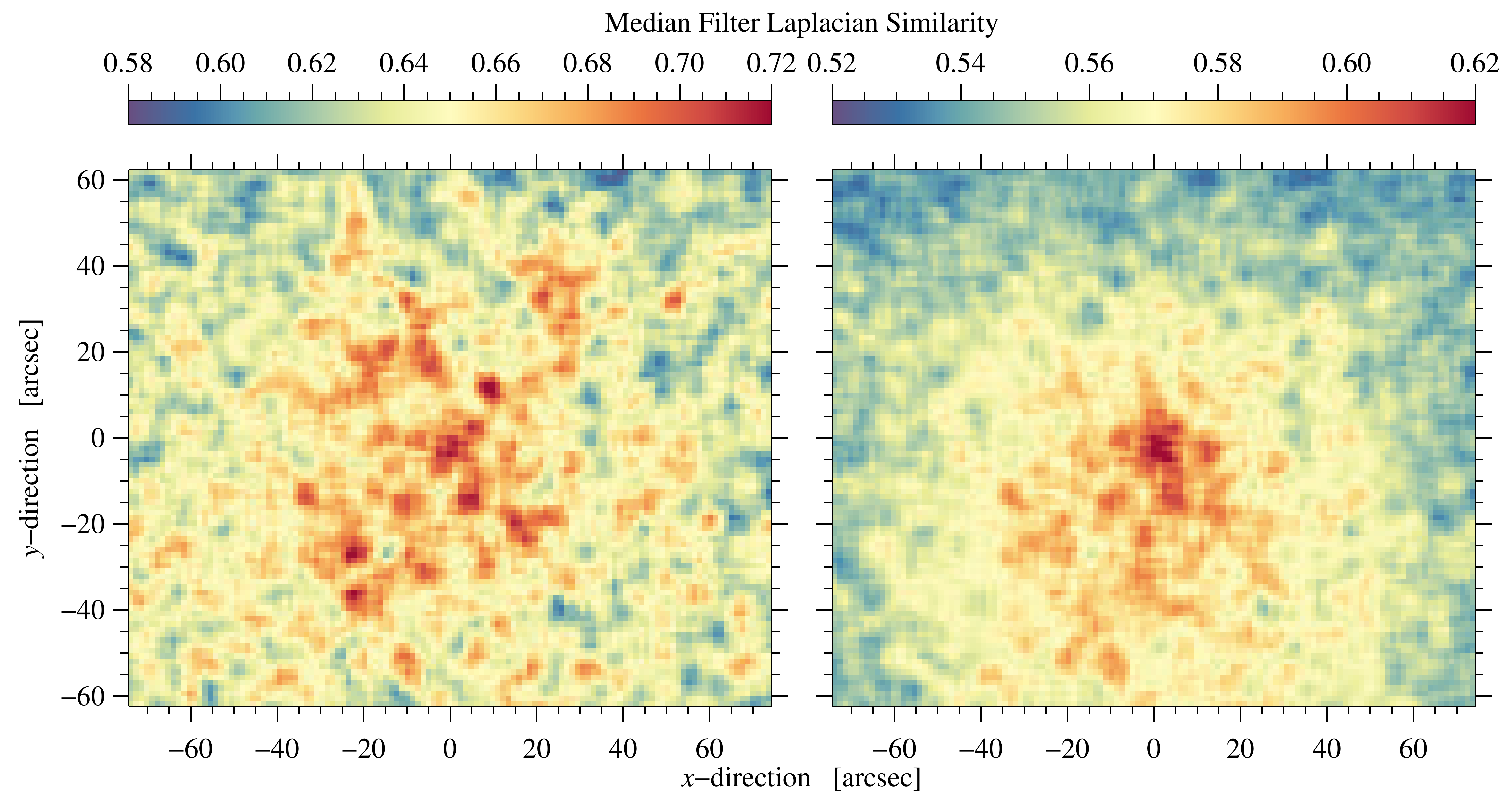}
\caption{Two-dimensional maps of the Median Filter Gradient Similarity
    (\textit{top}) and Median Filter Laplacian Similarity (\textit{bottom})
    corresponding to Figure~\ref{FIG01}. }
\label{FIG03}
\end{figure}

Two-dimensional false-color maps of MFGS and MFLS are compiled for the Fraunhofer G-band and the near-infrared $\lambda861.0$~nm in the top and bottom panels of Figure~\ref{FIG03}, respectively. The radial dependence of the image quality metrics is clearly evident in all panels. The region with the best AO correction is more compact in the infrared observations and has pronounced disk-like appearance. In contrast, the MFGS and MFLS values for the Fraunhofer G-band show more substructure, and clusters of mid-range values cover almost the entire FOV. The observations indicate again a significant correction by the AO system, well beyond the AO lock point and much larger than the isoplanatic patch. The MFGS and MFLS values cover very different ranges in the interval [0, 1]. The former metric is confined to a narrow interval below unity, whereas the latter metric stretches across a much larger interval of lower values. It is tempting to interpret this as a better performance of MFLS over MFGS. However, scatter plots of both metrics  exhibit a tight monotonic relationship among them (see Section~\ref{SEC06}). Thus, an advantage of MFLS is that image quality differences are better discernible.

High-contrast features, such as the bright rims of exploding granules \citep{Rast1995} or bright points, remain distinctly visible, even when computing image quality metrics over 80$\times$80-pixel regions. This leads to a ``blocky'' appearance of the two-dimensional maps, where these sampling windows slide across these features, with a step size of 20~pixels, which is four times smaller than the window size. The MFGS and MFLS methods are not very sensitive for images containing almost diffraction-limited information. This inability has its origin in the implementation of the MFGS and MFLS image quality metrics, which focus on 3$\times$3-pixel neighborhoods, thereby recognizing the presence of noise and discontinuous image information. Once the magnitude gradients and Laplacians become more and more similar $G(I_r) \approx G(I_m)$ and $L(I_r) \approx L(I_m)$, noise is effectively suppressed, and the image contains information of the observed object at high spatial frequencies.


\subsection{Differential Image Motion}

Besides rigid image displacement, anisoplanatism introduces differential image motion, which additionally degrades the spatial resolution and contrast of long-exposure images. The theory is well established \citep{Martin1987, Tokovinin2002} and led to a variety of instruments to quantify the seeing conditions, in particular the \textit{Solar Differential Image Motion Monitor} \citep[S-DIMM,][]{Beckers2001} for daytime seeing measurements. High-cadence, high-spatial resolution solar images deliver the field dependence of DIM \citep[\textit{e.g.},][]{Denker2007b, Denker2007a}. A two-dimensional correlation function is computed for each image in a stack of subfields with respect to the average image of the stack. The displacement of the maximum of the two-dimensional correlation function from the origin yields a displacement vector for each image in the stack. A robust DIM measure is derived by subtracting the mean $x$- and $y$-components from the vector of image displacements (removal of local tip-tilt component) and by taking the average of the magnitude of the displacement vectors. In principle, the term `residual' DIM should be used, referring to DIM after AO correction. `Normal' DIM values under daytime seeing conditions will be approached towards the periphery of the FOV, where the correction provided by the AO system becomes negligible. In the following the qualifier residual is dropped for sake of simplicity.

\begin{figure}[t]
\centering
\includegraphics[width=\textwidth]{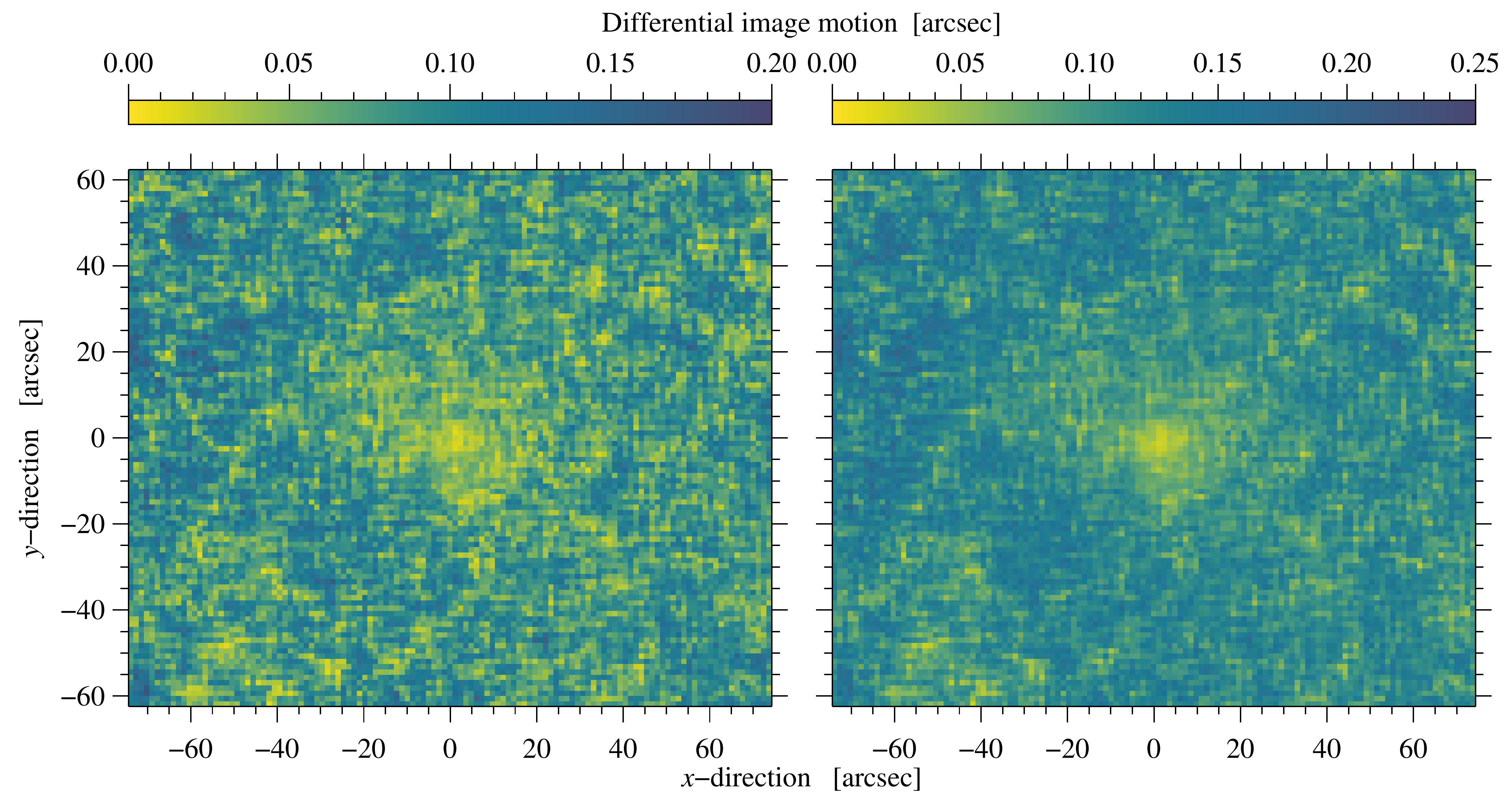}
\includegraphics[width=\textwidth]{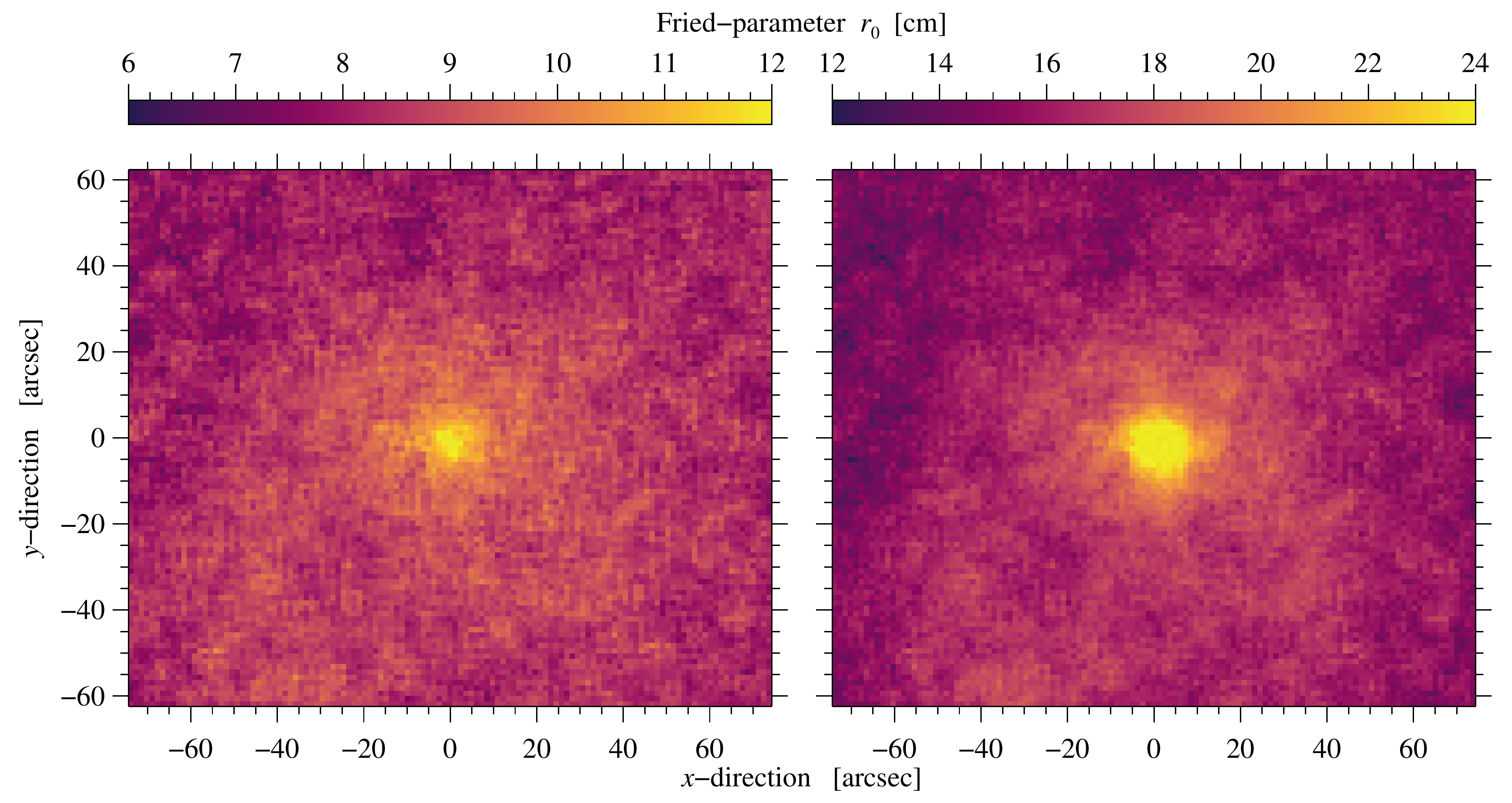}
\caption{Two-dimensional maps of the differential image motion (\textit{top}) 
    and Fried-parameter $r_0$ (\textit{bottom}) corresponding to Figure~\ref{FIG01}.}
\label{FIG04}
\end{figure}

In the upper panels of Figure~\ref{FIG04}, two-dimensional maps of the averaged DIM are shown. The maps are scaled in different ranges adapting to the different seeing conditions in the visible and near-infrared, \textit{i.e.}, 0.00\,--\,0.20~arcsec for Camera~1 and 0.00\,--\,0.25~arcsec for Camera~2, respectively. The two-dimensional maps of DIM are typically anti-correlated with the other image quality metrics and seeing parameters. The smallest DIM values are in the center of the FOV, where the AO correction is highest. Towards the periphery of the FOV, DIM values are a good representation of the seeing conditions of the free atmosphere. The DIM plays certainly an important role for the image quality of long-exposure images. However, it is the only parameter in this study, which is somewhat disconnect from the concept of image quality, when referring to short-exposure images, where the wavefront deformations are frozen.

\subsection{Fried-parameter}

The Fried-parameter $r_0$ \citep{Fried1974} was determined with the \textit{Spectral Ratio Technique} \citep[SRT,][]{vonderLuehe1984, vonderLuehe1987}, which requires that short-exposure images of an extended object such as the Sun contain spatial information down to the diffraction limit of the telescope. Applications of the technique to AO-corrected images are described in \citet{Denker2005b} and \citet{Puschmann2006a} for the speckle masking method \citep{Weigelt1983}. Starting point is the Fourier transform of the imaging equation \citep{vonderLuehe1984, vonderLuehe1987}
\begin{equation}
F(q) = F_0(q)\; S (q),
\label{EQN02 }
\end{equation}
where $F$ and $F_0$ are the Fourier transforms of the object's observed and true intensities, respectively, whereas $S$ denotes the optical transfer function (OTF). The two-dimensional spatial frequency $q$ was normalized by the cutoff frequency of the telescope $f_c = D / \lambda R$, where $\lambda$ is the wavelength of the incident light, $D$ the telescope diameter, and $R$ its focal length. The spectral ratio is defined as
\begin{equation}
\epsilon (q) = 
\frac{| \langle F_i (q) \rangle |^2}{\langle | F_i (q) |^2 \rangle} = 
\frac{| F_0 (q) |^2}{| F_0 (q) |^2}\;
\frac{| \langle S_i (q) \rangle |^2}{\langle | S_i (q) |^2 \rangle} =
\frac{| \langle S_i (q) \rangle |^2}{\langle | S_i (q) |^2 \rangle},
\label{EQN03}
\end{equation}
where (i) the object information $| F_0 (q) |^2$ cancels, (ii) the fraction on the left-hand side contains only observable quantities, \textit{i.e.}, the long-exposure $| \langle F_i (q) \rangle |^2$ and short-exposure $\langle | F_i (q) |^2 \rangle$ power spectra, and (iii) the fraction on the right-hand side can be modelled. The angle brackets $\langle \ldots \rangle$ in combination with the index $i$ denote an ensemble average. The long-exposure $| \langle S_i (q) \rangle |^2$ and short-exposure $\langle | S_i (q) |^2 \rangle$ transfer functions are based on the theory of wave propagation through a turbulent medium and can be described by a single parameter $r_0$ \citep{Fried1966, Korff1973}. The theory, on which these transfer functions are founded, does not include compensation of wavefront errors by AO systems. However, theoretical transfer functions still make it possible to derive a generalized Fried-parameter after partial wavefront compensation with an AO system \citep{Cagigal2000}.

Two-dimensional maps of the averaged Fried-parameter $r_0$ are presented in the lower panels of Figure~\ref{FIG04}. The highest $r_0$ values are concentrated in the central FOV within a radius of 5\,--\,10~arcsec. This core is surrounded by an halo that extends beyond the AO lock point by 20\,--\,40~arcsec. The display ranges differ by a factor of two taking into account the improved seeing conditions in the near-infrared, even though the improved seeing scales with $\lambda^{6/5}$. Since G-band data are available on all observing days, we included mean values and standard deviations of the Fried-parameter $r_0$ in Table~\ref{TAB01}. These values were computed from all two-dimensional maps of the Fried-parameter on a given observing day. The mean values do not reflect the Fried-parameters near the AO lock point, which are significantly higher. Nevertheless, the average values are appropriate to find the best observing day. The standard deviation indicates how the Fried-parameter changed during an observing day and how it varied within the observed FOV.

In general, the maps of the Fried-parameter appear much smoother than any of the other image quality metrics or seeing parameters. This is a direct effect of evaluating power spectra rather than contrasts features. The `polygonal' nature of solar granulation provides sufficient fine structure to recover diffraction-limited information, which is reflected in the radial profiles of the spectral ratio. Since the SRT analyzes the integral of the seeing fluctuations along the line of sight, whereas DIM is more closely bound to specific turbulence layers, disparities between Fried-parameter and DIM may hold clues about the height dependence of seeing.

%
%

\section{Radial Dependence of Adaptive Optics Correction}\label{SEC04}

\begin{figure}[t]
\centering
\includegraphics[width=\textwidth]{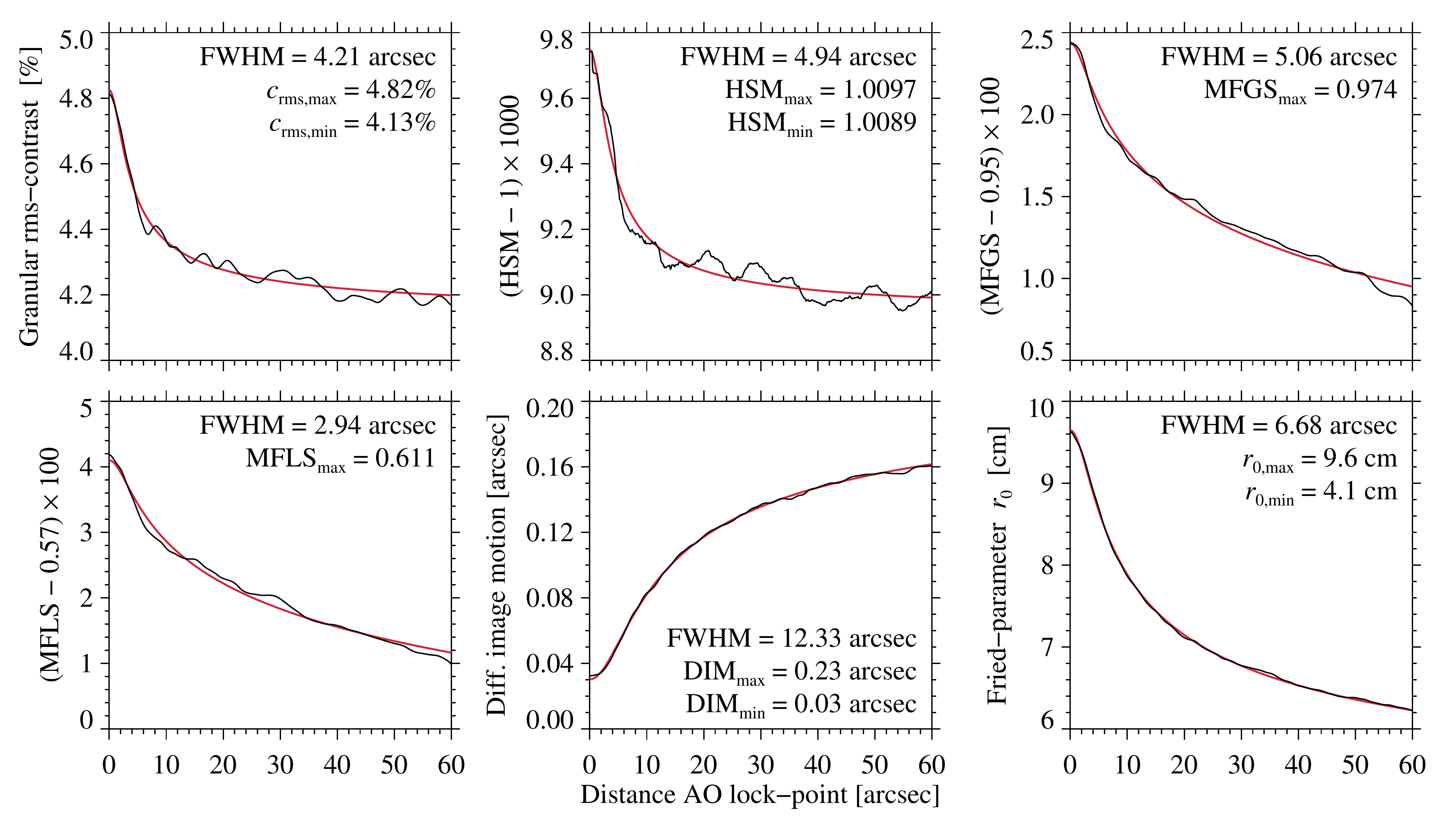}
\caption{Radial dependence of the average image quality metrics and seeing
     parameters with distance from the AO lock point (\textit{black}) for all G-band datasets. All relationships are well represented by Moffat functions (\textit{red}) with a constant term. The FWHM of the Moffat fits are given in the corner of the panels, along with the maximum and minimum values if reasonable. }
\label{FIG05}
\end{figure}

The two-dimensional maps of image quality metrics and seeing parameters presented in Section~\ref{SEC03} show clear evidence that the AO correction extends well beyond the isoplanatic patch centered on the lock point of the AO system. To quantify the radial dependence of the AO correction with distance from the lock point, all G-band $\lambda$430.7~nm data were used to create average maps of the image quality metrics and seeing parameters. The $121 \times 101$ grid points were resampled to the size of the input images with $2480 \times 2080$ pixels using cubic convolution interpolation. In the next step, azimuthal averages were computed to determine the radial dependence of the image quality metrics and seeing parameters. The MPFIT software package \citep{Markwardt2009} provides many functions and tools for non-linear least-squares fitting, including the Moffat function, which is often a better choice for modeling PSFs than Gaussian functions \citep{Moffat1969}. The radial dependence of the image quality metrics and seeing parameters is summarized in Figure~\ref{FIG05}, which includes the average observed profiles and a fit with the Moffat function including its FWHM. The quality of the fits is very good but converges to negative values for MFGS and MFLS for infinite distances from the AO lock point, which is not physically meaningful. Thus, minima and maxima of the metrics and parameters are only presented in the corner of the panels if reasonable.

The radial profiles for granular rms-contrast and Helmli-Scherer means look very similar, a narrow central core and a shallow halo. The profile still shows some substructure beyond the central core, which seems to result from contrast features that survived the averaging process of 159 datasets. The FWHM of 4.21~arcsec and 4.94~arcsec are in both cases comparable. However, the width of the central core can be much larger for individual datasets (see Section~\ref{SEC03}). The radial MFGS and MFLS profiles appear visually the same but the width of the central core differs significantly, \textit{i.e.}, 5.06~arcsec \textit{vs.} 2.94~arcsec, respectively. This can be attributed to the extended halo in the radial profiles, which only shows minor substructure. In contrast, the radial profiles of differential image motion and Fried-parameter $r_0$ are very smooth and do not differ much from the respective fits with Moffat functions. Their FWHM of 12.33~arcsec and 6.68~arcsec are significantly larger compared to all other image quality metrics. They also exhibit extended halos, which significantly aids \textit{post facto} image restoration. Inspecting the extreme values at infinite distance from the AO lock point supports this claim.

\begin{figure}[t]
\centering
\includegraphics[width=\textwidth]{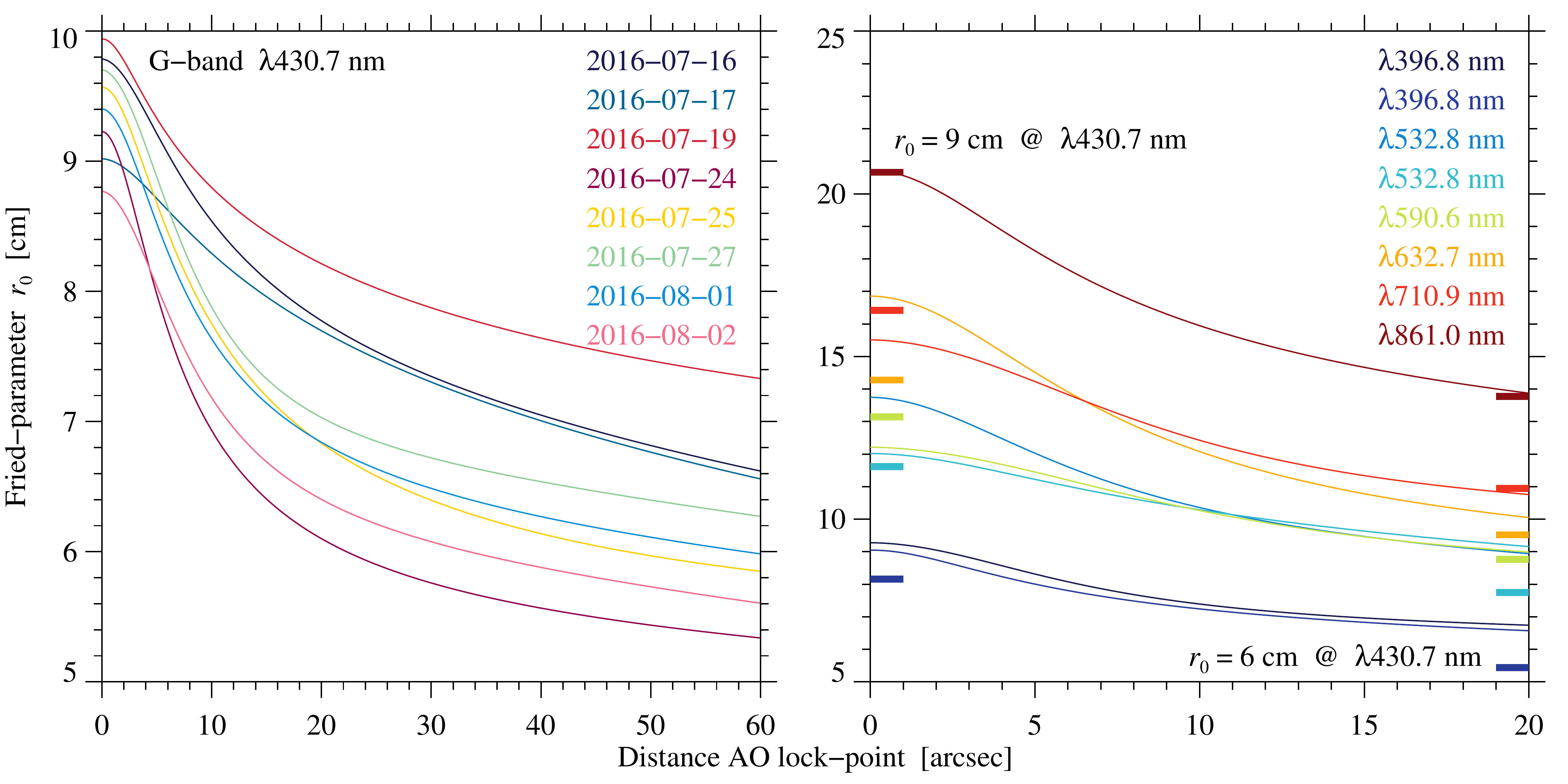}
\caption{Radial dependence of the average Fried-parameter $r_0$ with distance 
    from the AO lock point for the daily G-band datasets (\textit{left}). Only the Moffat fits to the radial profiles are displayed for clarity. Radial dependence of the average Fried-parameter $r_0$ with distance from the AO lock point for the specified wavelengths (\textit{right}). The radial profiles were corrected for the daily variation of seeing quality based on the halos of the G-band profiles but only the inner part of the profile is shown. Assuming that the Fried-parameters $r_0 = 9$ and 6~cm at $\lambda$430.7~nm scales with $\lambda^{6/5}$, the color-coded short horizontal bars on the left and right ordinates indicate the Fried-parameter $r_0$ for the wavelengths specified in the upper-right corner.}
\label{FIG06}
\end{figure}

The left panel of Figure~\ref{FIG06} is the pendant to the panel with the Fried-parameter $r_0$ in Figure~\ref{FIG05} but it shows the average radial profiles for each observing day. The shape is overall very similar with a distinct core-halo pattern. However, 2016 July~17 and~24 display higher and lower core-halo ratios, indicating different types of seeing conditions. Short exposure images on 2016 July~17 are characterized by a more uniform image quality across the FOV, while those on 2016 July~24 give the impression of looking through a keyhole, where only the central part of the FOV appears sharp.

The right panel of Figure~\ref{FIG06} is an attempt to validate the $\lambda^{6/5}$ wavelength dependence for the radial profiles of the Fried-parameter $r_0$. The radial profiles for the observed wavelength are of course influenced by the varying seeing conditions. Therefore, the daily radial $r_0$ profiles for the G-band data were used for calibration, assuming that the free atmosphere can be characterized by $r_0$ values far away from the AO lock point. Core and halo of the $\lambda$861.0~nm profile corresponds to Fried-parameters $r_0 = 9$ and 6~cm at $\lambda$430.7~nm (see lower panels of Figure~\ref{FIG04} for a two-dimensional representation of the core-halo structure). Thus, these values were chosen as reference. The respective $r_0$ values for the other observed wavelength are marked on the ordinates of the plot panel. However, only a qualitative agreement could be reached. The varying core-halo ratios on different observing days obscure a clear trend. Even intersecting radial profiles are seen for the $\lambda$632.7~nm and $\lambda$710.9~nm data.

%
%

\section{Correlation Analysis of Wavelength Dependence}\label{SEC05}

\begin{figure}[p]
\centering
\includegraphics[width=0.93\textwidth]{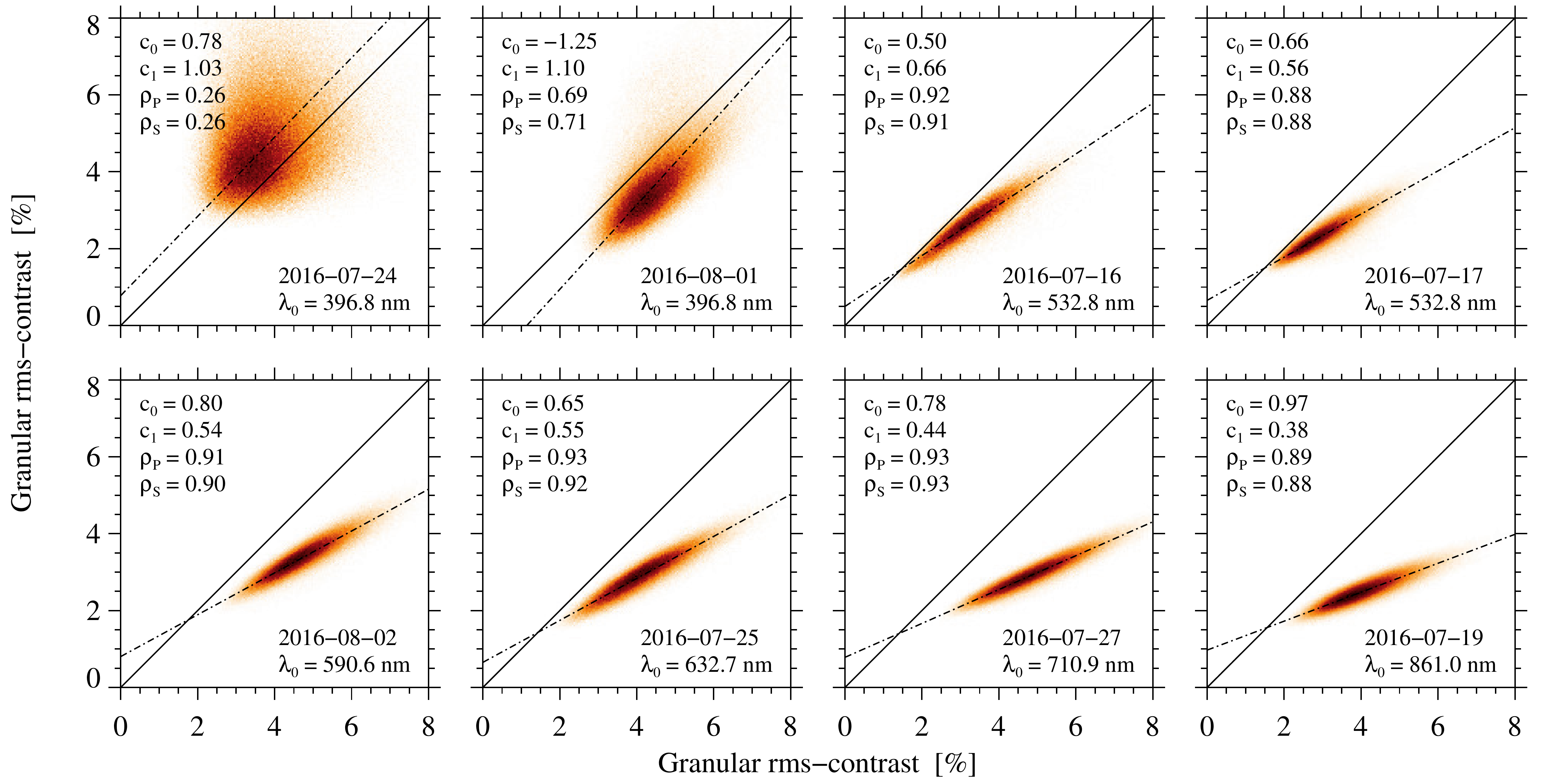}\vspace*{3pt}\\
\includegraphics[width=0.93\textwidth]{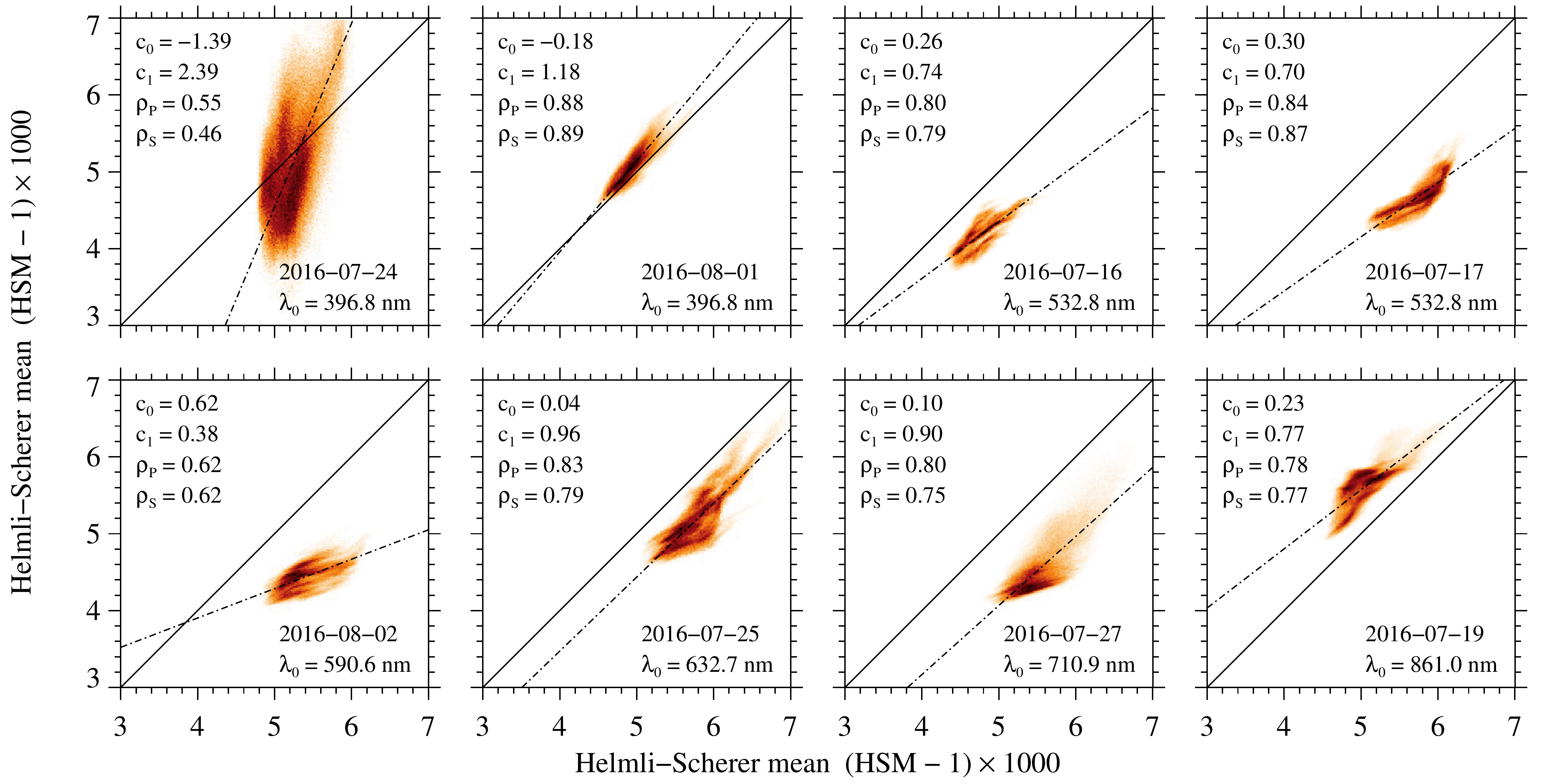}\vspace*{3pt}\\
\includegraphics[width=0.93\textwidth]{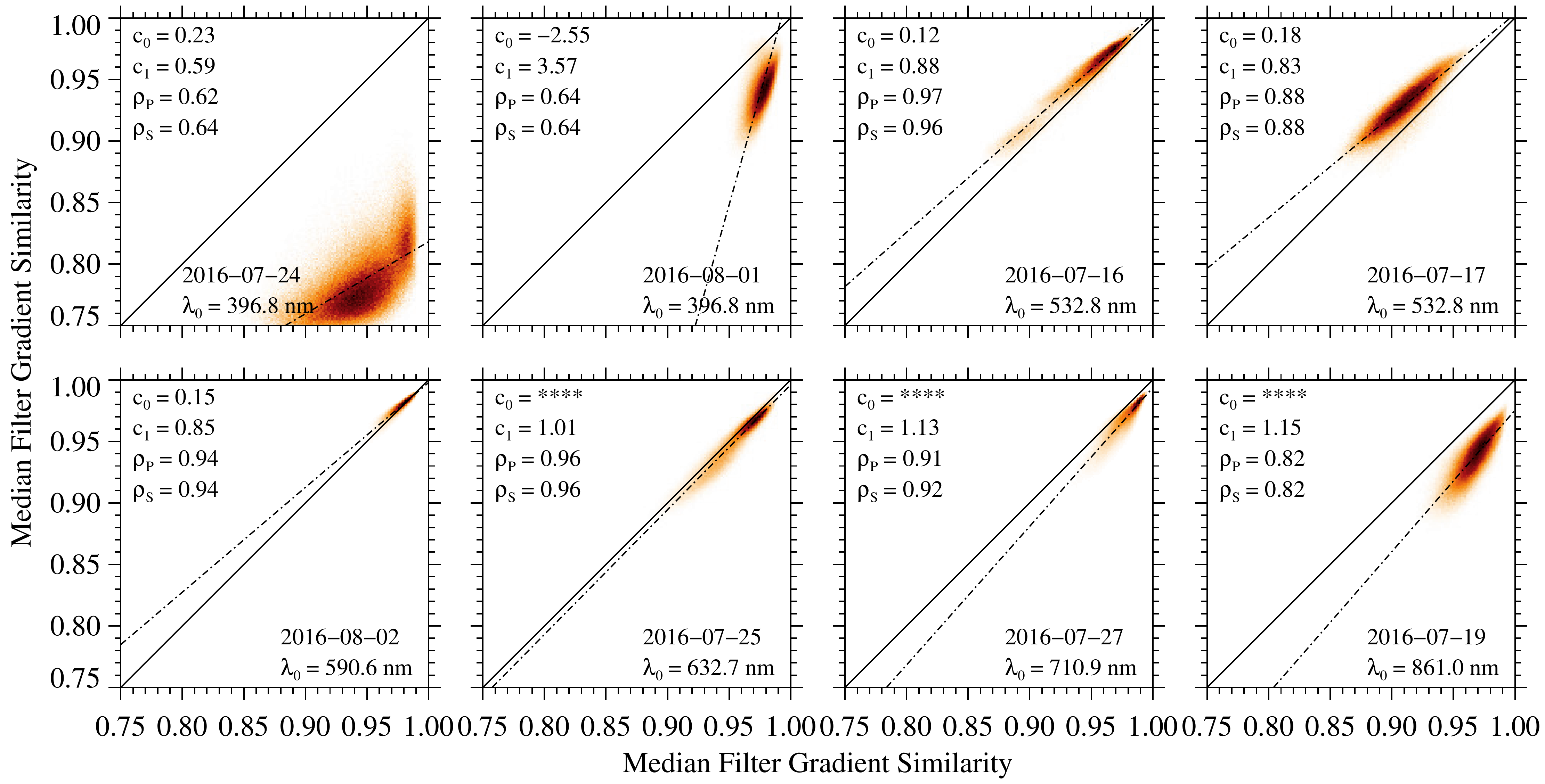}
\caption{Two-dimensional histograms of granular rms-contrast (\textit{top}),
    Helmli-Scherer Mean (\textit{middle}), and Median Filter Gradient Similarity  (\textit{bottom}) as a function of wavelength (\textit{ordinate}). The G-band values (\textit{abscissa}) serve as reference. The first panels refer to the narrow-band Ca\,\textsc{ii}\,H images. The identity line (\textit{solid}) facilitates a better comparison with a linear model (\textit{dashed}), where $c_0$ and $c_1$ are vertical intercept and gradient, respectively. The suitability of the model can be deduced from Pearson's linear and Spearman's rank-order correlation coefficients $\rho_\mathrm{P}$ and $\rho_\mathrm{S}$, respectively.}
\label{fig07}
\end{figure}

\begin{figure}[p]
\centering
\includegraphics[width=0.93\textwidth]{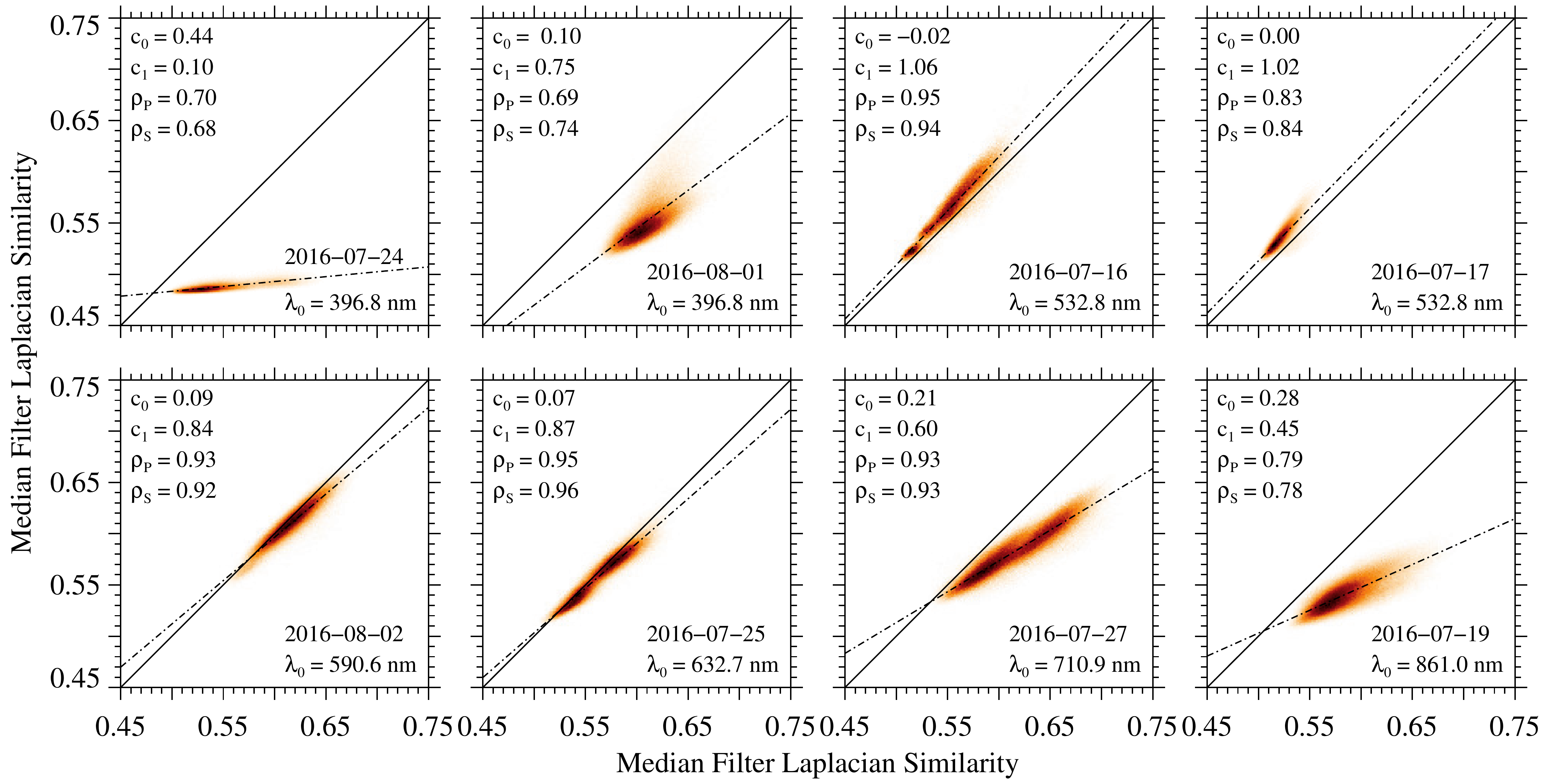}\vspace*{3pt}\\
\includegraphics[width=0.93\textwidth]{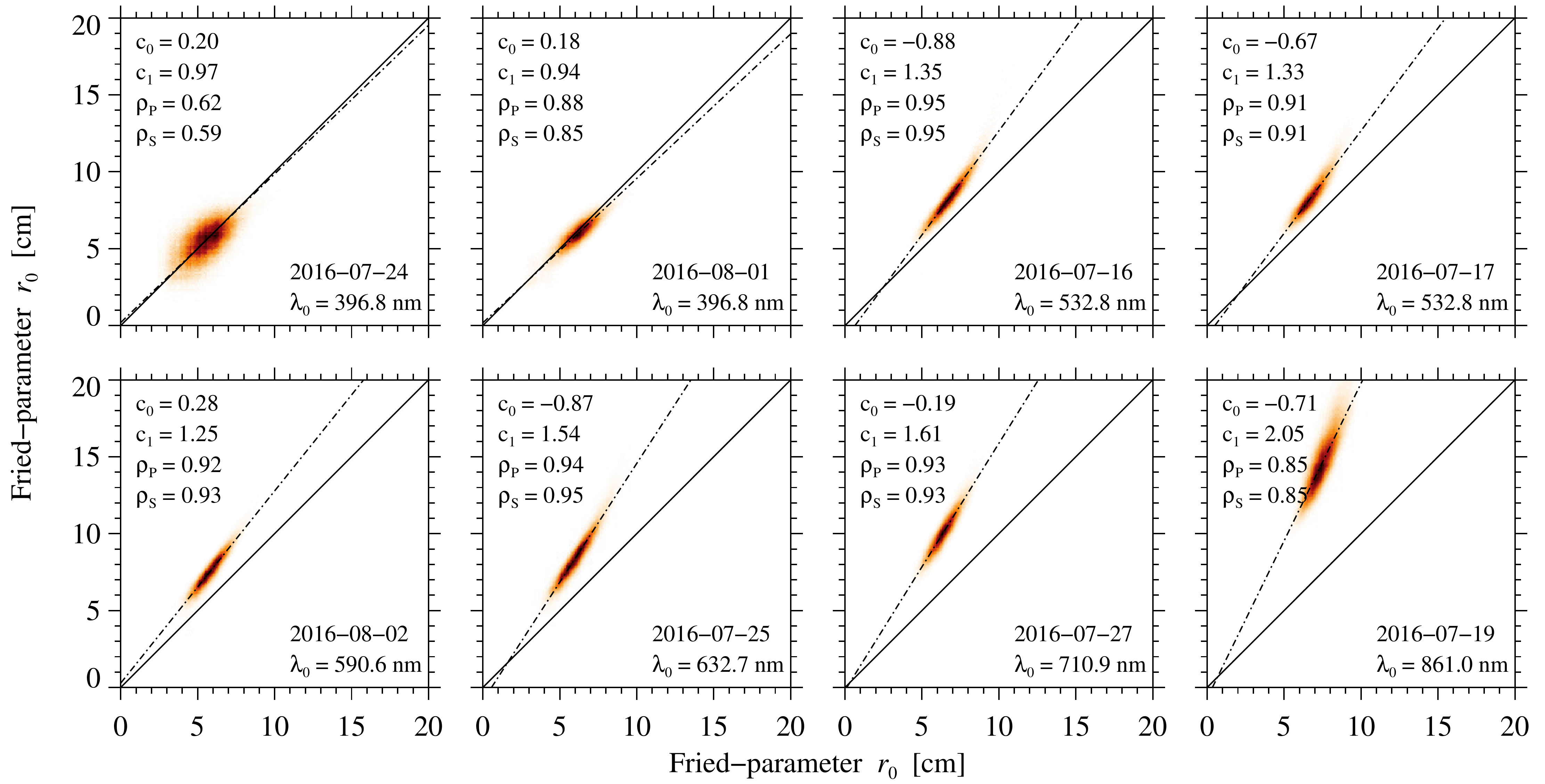}\vspace*{3pt}\\
\includegraphics[width=0.93\textwidth]{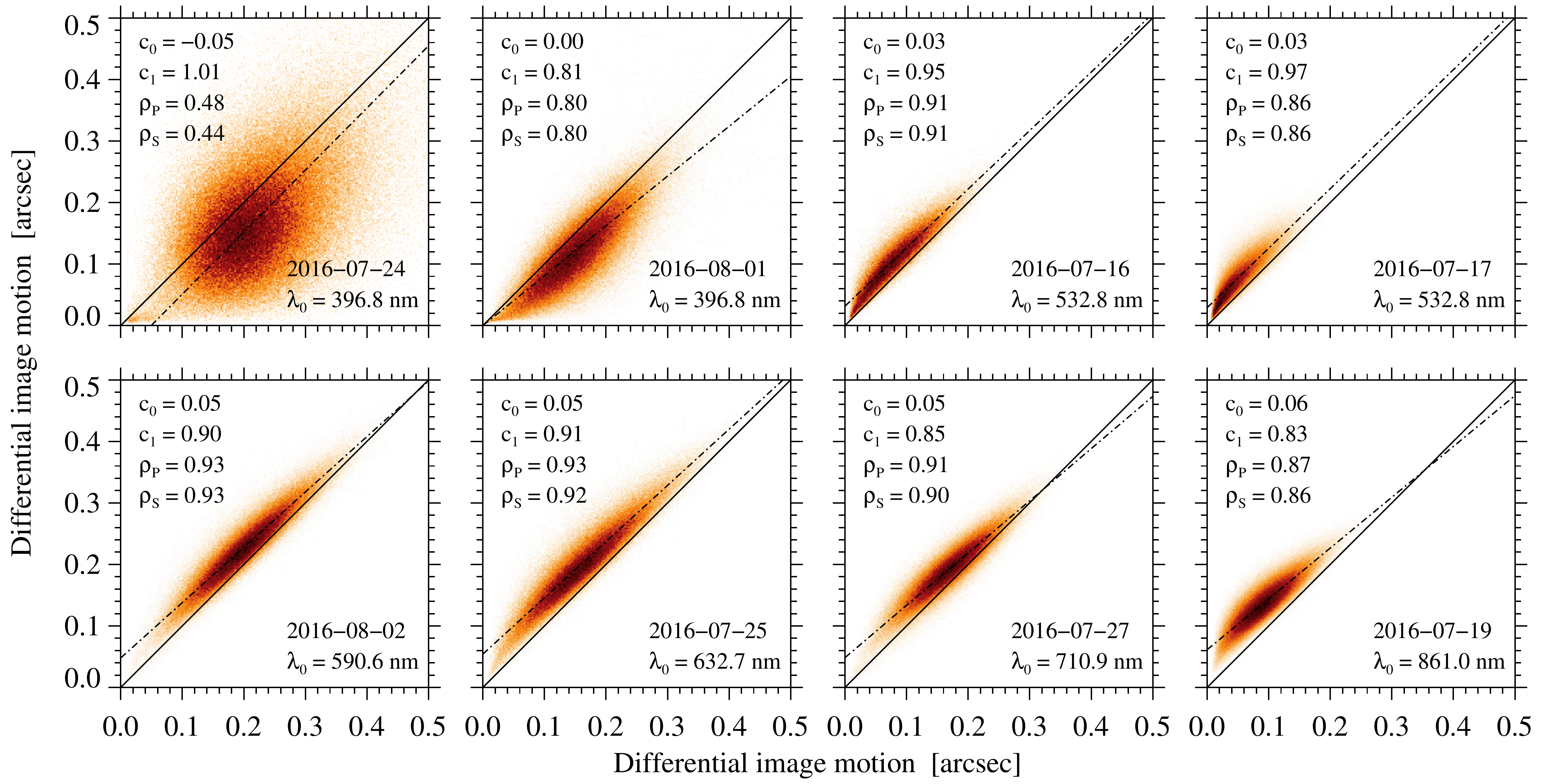}
\caption{Two-dimensional histograms of Median Filter Laplacian Similarity
    (\textit{top}), Fried-parameter $r_0$  (\textit{middle}), and differential image motion  (\textit{bottom}) as a function of wavelength (\textit{ordinate}). The G-band values (\textit{abscissa}) serve as reference. The first panels refer to the narrow-band Ca\,\textsc{ii}\,H images. The identity line (\textit{solid}) facilitates a better comparison with a linear model (\textit{dashed}), where $c_0$ and $c_1$ are vertical intercept and gradient, respectively. The suitability of the model can be deduced from Pearson's linear and Spearman's rank-order correlation coefficients $\rho_\mathrm{P}$ and $\rho_\mathrm{S}$, respectively.}
\label{fig08}
\end{figure}

The number of datasets per observing day at disk center is summarized in Table~\ref{TAB01}. After careful alignment, 8\,--\,32 datasets per day with $121 \times 101$ overlapping isoplanatic patches are available to determine image quality metrics and seeing parameters and their wavelength dependence using the G-band $\lambda$430.7~nm observations as a reference. Simple scatter plots will contain up to a few hundred thousand data points, which will crowd the display. Thus, two-dimensional histograms with $200 \times 200$ bins are more appropriate to capture the wavelength dependence of metrics and parameters. The results are depicted in Figures~\ref{fig07} and~\ref{fig08} in the order of appearance in Section~\ref{SEC03}. The histograms are scaled linearly between zero and maximum number of occurrence. Empty bins are displayed in white and darker regions refer to a higher frequency of occurrence. The metrics and parameters are then sorted in the sub-figures according to wavelength, which is not the chronological order. The first panel refers to the narrow-band, line-core Ca\,\textsc{ii}\,H images. The plotting range is the same on both axes to facilitate better comparison of the histograms. Narrow- and broad-band Ca\,\textsc{ii}\,H images and G-band images contain bright points and varying contributions of the chromospheric network. The morphological differences for Ca\,\textsc{ii}\,H line-core images are significant in comparison to other broad-band images. However, the impact of small-scale brightenings progressively diminishes from broad-band Ca\,\textsc{ii}\,H to G-band images. In general, these differences in morphology are important when interpreting the two-dimensional histograms. 

In many cases, a linear model is well suited to describe the relationship between values at two different wavelengths. The fit coefficients $c_0$ and $c_1$ for the vertical intercept and gradient, respectively, were determined by minimizing the $\chi^2$-error statistic between observations and model. In some cases, a more robust line fit was required for the Ca\,\textsc{ii}\,H data, when the uncertainties along abscissa and ordinate were comparable \citep[e.g.,][]{Street1988}. The linear fit and the identity line are given in each panel to clarify trends. Pearson's linear and Spearman's rank-order correlation coefficients $\rho_\mathrm{P}$ and $\rho_\mathrm{S}$, respectively, are good indicators if the linear model matches the observed data. The fit and correlation coefficients are given in each plot panel for reference.

The two-dimensional histograms in the top panels of Figure~\ref{fig07} show a linear relationship for granular rms-contrast in the G-band with all other wavelengths. However, the correlation coefficients are significantly lower for the narrow- and broad-band Ca\,\textsc{ii}\,H images, where also the linear fit is shifted parallel to the line of equality. In the following, the  Ca\,\textsc{ii}\,H data are excluded from the discussion. In all other cases, the fit intercepts this line at almost the same locations, and the gradients of the fits decrease with increasing wavelength. In general, the two-dimensional histograms appear compact without any substructures. This is not the case for the histograms of the HSM, which exhibit sub-structure and clusters with a higher frequency of occurrence. In addition, the correlation coefficients are lower compared to the granular rms-contrast, and a clear relationship is absent between the linear fit and the line of equality.

The correlation coefficients of granular rms-contrast and MFGS/MFLS are comparable (see bottom and top panels of Figures~\ref{fig07} and~\ref{fig08}, respectively). The gradient of the MFGS is increasing with wavelength while that of the MFLS is decreasing. In addition, the intersection point of the linear fit is located at unity in the upper right corner of the plot panels, whereas the MFLS intersection point is situated in proximity to the origin in the lower left corner of the plot panels. Some of the MFLS histograms show a bimodal structure. This bimodal structure maybe related to changing seeing conditions during the observing day, whereas the much more detailed sub-structure encountered in the histograms of the HSM is presumably due to the presence of high-contrast features in the observed FOV.

Fried-parameter $r_0$ and differential image motion (see middle and bottom panels of Figure~\ref{fig08}) trace their origin back to image restoration. Both parameters require time-series data and can be computed for regions with the size of the isoplanatic patch. Their two dimensional histograms are compact and do not show any substructure. In most cases, the correlation coefficients of the Fried-parameter $r_0$ are slightly higher compared to the differential image motion. The gradient of the linear fit increases with wavelength for the Fried-parameter $r_0$, whereas the gradient remains close the line of equality for the differential image motion.

In summary, the correlation analysis of the wavelength dependence of the image quality metrics and seeing parameters reveals that linear models represent the two-dimensional histograms well -- with the notable exception of the HSM. In addition, the morphology of chromospheric fine structure in narrow- and broad-band Ca\,\textsc{ii}\,H data differs too much from photospheric quasi-continuum and G-band images so that a direct comparison is not meaningful. The two-dimensional histograms did not take into account the location of the isoplanatic patches within the FOV, \textit{i.e.}, changes of the seeing conditions, in particular changing contributions from layers within Earth's atmosphere, have the potential to change the relative contributions with distance from the lock point of the AO system. However, the compactness of the two-dimensional histograms, the good performance of the linear models, and the trends of the gradients argue for a monotonic wavelength dependence of the image quality metrics and seeing parameters.

%
%

\section{Correlation between Image Quality Metrics and Seeing Parameters}
\label{SEC06}

\begin{figure}
\centering
\includegraphics[width=1.\textwidth]{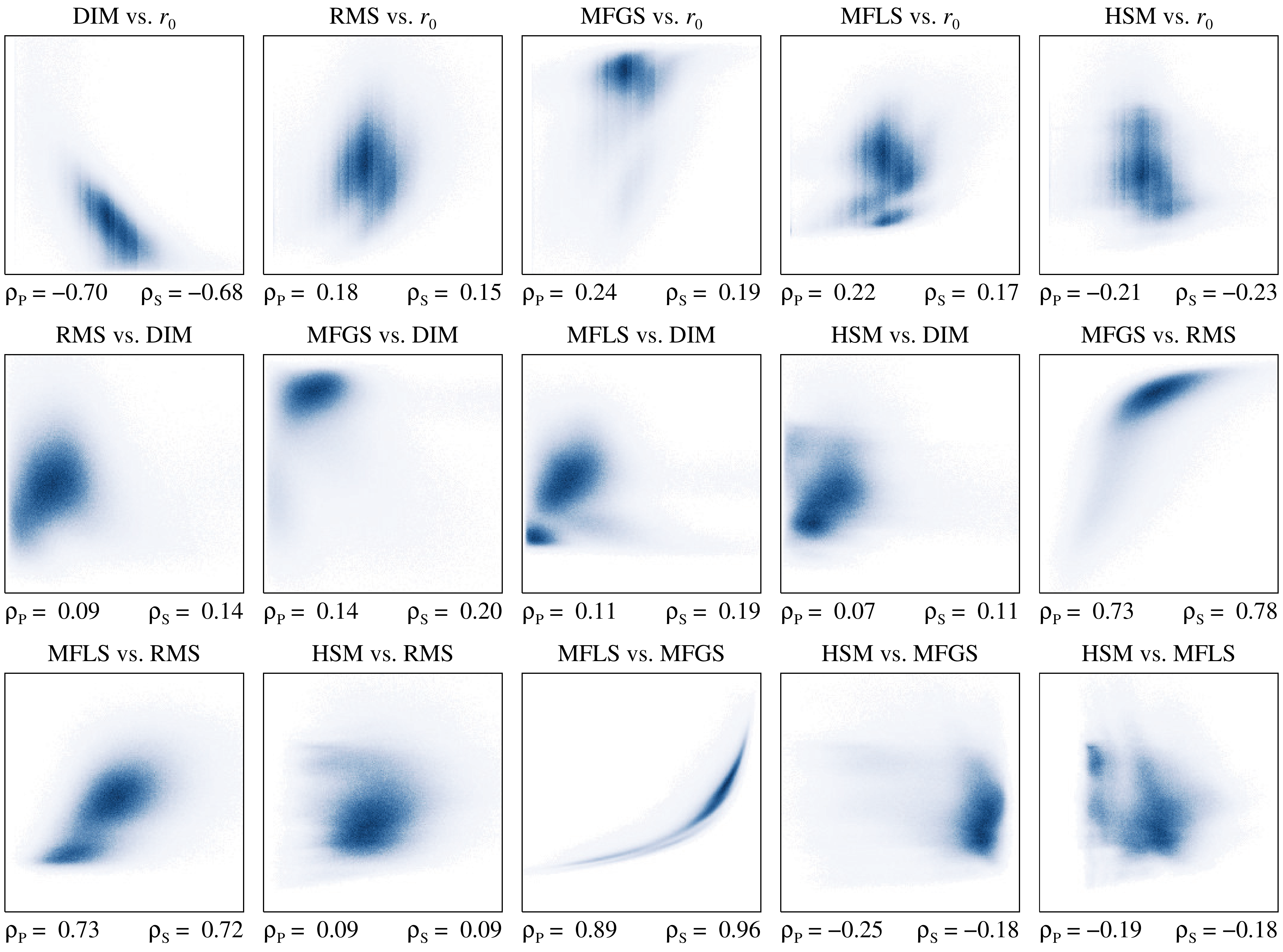}\smallskip
\caption{Scatter plots implemented as two-dimensional histograms showing the 
    correlation between image quality metrics and seeing parameters, which are displayed in the intervals RMS $\in$ [1.0\%, 8.0\%], HSM $\in$ [1.004, 1.007], MFGS $\in$ [0.85, 1.0], MFLS $\in$ [0.45, 0.75], DIM $\in$ [0.0~arcsec, 0.7~arcsec], and $r_0$ $\in$ [2~cm, 12~cm]. Pearson's linear and Spearman's rank-order correlation coefficients $\rho_\mathrm{P}$ and $\rho_\mathrm{S}$, respectively, quantify how well a linear or monotonic function describes the relationship between two parameters.}
\label{FIG09}
\end{figure}

Six image quality metrics and seeing parameters are included in this study. Thus, 15 pairwise comparisons are possible to investigate multicollinearity and to assess the predictive power of individual metrics and parameters. The G-band observations offer the best dataset for such an investigation because they encompass all observing days with different seeing conditions. The scatter plots in Figure~\ref{FIG09} are implemented as two-dimensional histograms with $200 \times 200$ bins for the pairwise comparison of metrics and parameters, which are given in the titles of the plot panels. Darker blue colors indicate a higher frequency of occurrence. The axis labels are omitted to prevent crowding the plot but the ranges for ordinates and abscissae are given as intervals in the caption of Figure~\ref{FIG09}. The intervals are narrower (with the exception of the Helmi-Scherer mean) than in Figures~\ref{fig07} and~\ref{fig08} because only a single wavelength setting is analyzed. The two-dimensional maps of image quality metrics and seeing parameters (see Section~\ref{SEC03}) contain sometimes erroneous values, in particular at the peripheral FOV and under mediocre seeing conditions. These values are removed based on the difference between the original map and a median-filtered map (3$\times$3 neighborhood), using a threshold of three times the standard deviation of the difference map. Erroneous values flagged in one map are also removed from the remaining five maps. Note that the location of isoplanatic patches within the FOV is scrambled in the two-dimensional histograms (Fig.~\ref{FIG09}).

Differential image motion and Fried-parameter $r_0$ are clearly anticorrelated as expected for these seeing parameters. A strictly monotonous dependence is clearly evident for the gradient- and curvature-based image quality metrics, \textit{i.e.}, MFLS \textit{vs.} MFGS, where the rank-order correlation coefficient $\rho_\mathrm{S}$ is almost unity. A similar behavior was already observed for different MFGS implementations \citep{Denker2018b}. Other well-defined correlations include granular rms-contrast with MFGS and MFLS, respectively, whereas the former shows some hysteresis under mediocre seeing conditions and the latter exhibits a bimodal distribution. Such distributions are frequently encountered, \textit{e.g.}, for MFLS \textit{vs.} DIM and HSM \textit{vs.} DIM. These are first indications that varying seeing conditions, \textit{i.e.}, varying contributions from ground-layer seeing and those from the free atmosphere, lead to more structured two-dimensional histograms. However, it is not obvious how to connect the image quality metrics and seeing parameters to the height dependence of the refractive index structure function $C_N^2(h)$ commonly used in seeing measurements. Surprisingly, the correlation between Fried-parameter $r_0$ and granular rms-contrast is low, even though the image contrast is often used as criterion for the prevailing seeing conditions. The other notable finding is that the Helmli-Scherer mean shows weak positive and negative correlations with all other metrics and parameters. In summary, apparent collinearity is observed for the image quality metrics granular rms-contrast, MFGS, and MFLS and for the seeing parameters differential image motion and Fried-parameter $r_0$. 

\begin{figure}[t]
\centering
\includegraphics[width=1.\textwidth]{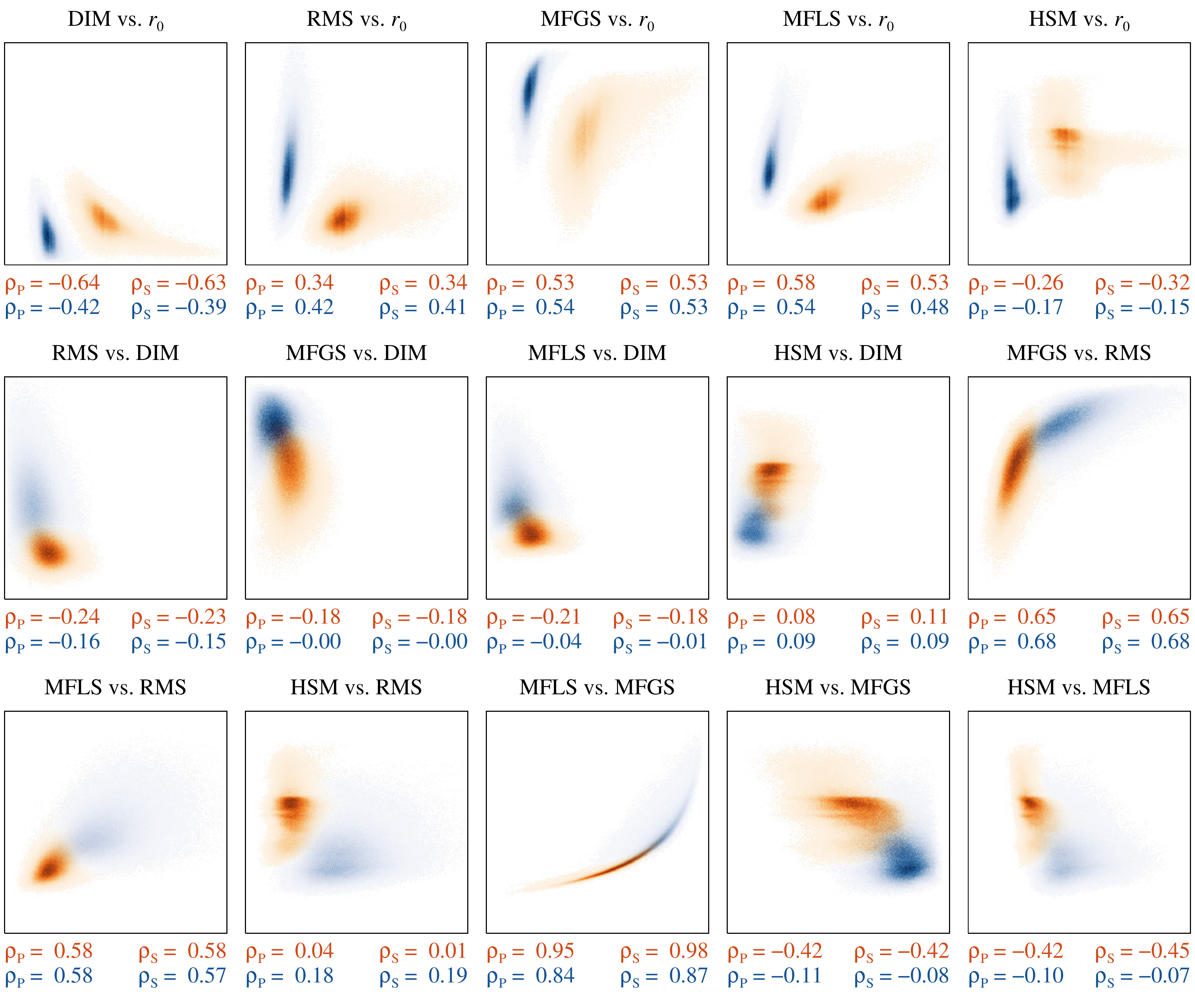}\smallskip
\caption{Scatter plots implemented as two-dimensional histograms showing the
    correlation between image quality metrics and seeing parameters for the wavelength settings $\lambda 861.0$~nm (\textit{red}) and G-band (\textit{blue}), which are displayed in the intervals RMS $\in$ [1.0\%, 8.0\%], HSM $\in$ [1.004, 1.007], MFGS $\in$ [0.85, 1.0], MFLS $\in$ [0.45, 0.75], DIM $\in$ [0.0~arcsec, 0.7~arcsec], and $r_0$ $\in$ [2~cm, 30~cm]. Pearson's linear and Spearman's rank-order correlation coefficients $\rho_\mathrm{P}$ and $\rho_\mathrm{S}$, respectively, quantify how well a linear or monotonic function describes the relationship between two parameters.}
\label{FIG10} 
\end{figure}

The data recorded on July~19 includes the most sets per day and was taken under the best seeing conditions (Table~\ref{TAB01}). In addition, G-band $\lambda$430.7~nm and $\lambda$861.0~nm observations have the largest separation in wavelength. Therefore, a similar correlation analysis as presented in Figure~\ref{FIG09} will be carried out for two wavelength settings in Figure~\ref{FIG10}. The results of the $\lambda$861.0~nm observations are displayed using a red color scale, while the blue color scale represents the G-band $\lambda$430.7~nm data. Doubling the wavelength leads to a clear separation of the two-dimensional frequency distributions. Since the seeing improves when observing at longer wavelengths, the results are displayed across larger intervals listed in the caption of Figure~\ref{FIG10}. The correlation coefficients $\rho_\mathrm{P}$ and $\rho_\mathrm{S}$ are very similar and only show slight variations between the two observed wavelengths. Correlation or anticorrelation is easily deduced from the scatter plots by visually connecting the centers of gravity of the two frequency distributions. The shape of the distributions often implies a linear relationship but in some cases, \textit{e.g.}, MFGS \textit{vs.} $r_0$, a curvature is evident. The curved distributions for MFLS \textit{vs.} MFGS are very narrow and merge smoothly. The main weakness of MFGS and MFLS is that they start to saturate both under very poor seeing conditions and when the seeing is extremely good. Therefore, they are not very sensitive to changes in the quality of superb and almost diffraction-limited images \citep[see][]{Denker2018b}. The negative correlation between differential image motion and other seeing parameters and image quality metrics is expected. However, the Helmli-Scherer mean shows a similar trend, which is unexpected because \citet{Popowicz2017} reported a positive correlation with improving seeing and attested a good performance as an image quality metric. The width of the frequency distributions involving the Helmli-Scherer mean is also comparatively large and often includes a halo. The granular rms-contrast is often used as a proxy of the seeing quality. Such a superior performance is, however, not evident in the scatter plots. Here, the contrast is more closely tied to the MFGS and MFLS metrics.

\begin{figure}
\centering
\includegraphics[width=0.6\textwidth]{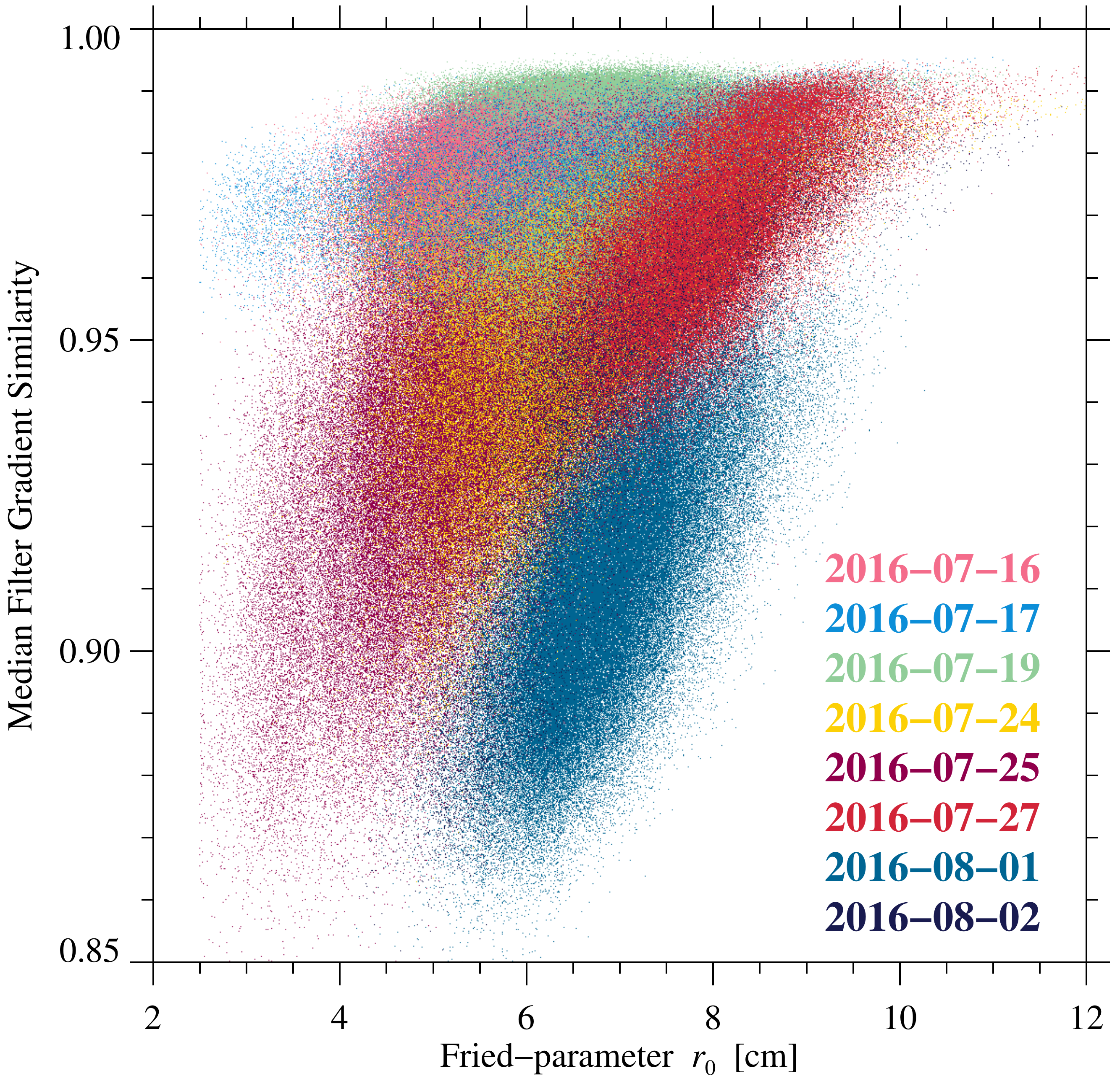}\smallskip
\caption{Scatter plot showing the correlation between the seeing parameter
    r$_0$ $\in$ [2, 12] and the image quality metric MFGS $\in$ [0.85, 1.0]. The color code refers to the observing date given in the lower right corner.}
\label{FIG11}
\end{figure}

Different seeing conditions may affect shape and structure of the two-dimensional histograms. Based on the histogram MFGS \textit{vs.} Fried-parameter $r_0$ (middle panel in the top row of Figure~\ref{FIG09}), the scatter plot in Figure~\ref{FIG11} depicts how the seeing on different days contributes to the overall distribution. This scatter plot comprises more than 1.5~million data points referring to individual isoplanatic patches, which are plotted from left to right sorted by increasing Fried-parameter $r_0$. Just plotting the data points day by day will selectively cover up some of the daily variations. This plot should not be interpreted as a two-dimensional histogram since certain locations in the scatter plot are overplotted multiple times. Color coding the data points reveals that the daily observations are strongly clustered, sometimes exhibiting clearly linear relationships, while at other times \textit{e.g.,} on 2016 July~25, the MFGS seems to be independent of the  Fried-parameter $r_0$. These daily variations raise the question if a systematic relationship exists among seeing parameters and image quality metrics, which contains information about contributions to seeing and the height dependence of seeing. This question motivated the following section, where an unsupervised machine learning algorithm is employed to visualize relationships between various parameters and metrics including their daily variations.

%
%

\section{Classification Using Uniform Manifold Approximation and Projection}\label{SEC07}

Different image quality metrics and seeing parameters may be sensitive to different seeing properties encoded in the time-series data. Since the time-series are comparatively short, it is a reasonable assumption that the seeing conditions, \textit{i.e.}, the relative contribution of different atmospheric layers to the seeing, did not change much during the short observing period on a given observing day. Even though, more instantaneous fluctuations of the seeing are still present in the two-dimensional maps of the image quality metrics and seeing parameters. \textit{Uniform Manifold Approximation and Projection} \citep[UMAP,][]{McInnes2018} is a good choice to explore the systematic relations among the metrics and parameters. In addition, UMAP can be used to find the best combination and minimal number of metrics and parameters, which explains the observed seeing variation. UMAP is an unsupervised machine learning algorithm, which provides non-linear dimensionality reduction and furnishes visualization of patterns or clustering in data without having any prior knowledge of the labels (\textit{i.e.}, in our case the observing dates) describing the data. The underlying theoretical idea of UMAP is largely based in manifold theory and topological data analysis. UMAP belongs to the class of k-neighbour based graph learning algorithms similar to other unsupervised algorithm such as \textit{t-distributed Stochastic Neighbor Embedding} \citep[t-SNE,][]{VanderMaaten2008}. However, UMAP's scaling performance is often better compared to t-SNE. Primarily designed to preserve local structures, UMAP also effectively preserves structure on global scales as discussed in \citet{McInnes2018} who provide details on the UMAP implementation and compare it to other algorithms.

\begin{figure}[t]
\centering
\includegraphics[width=1.\textwidth]{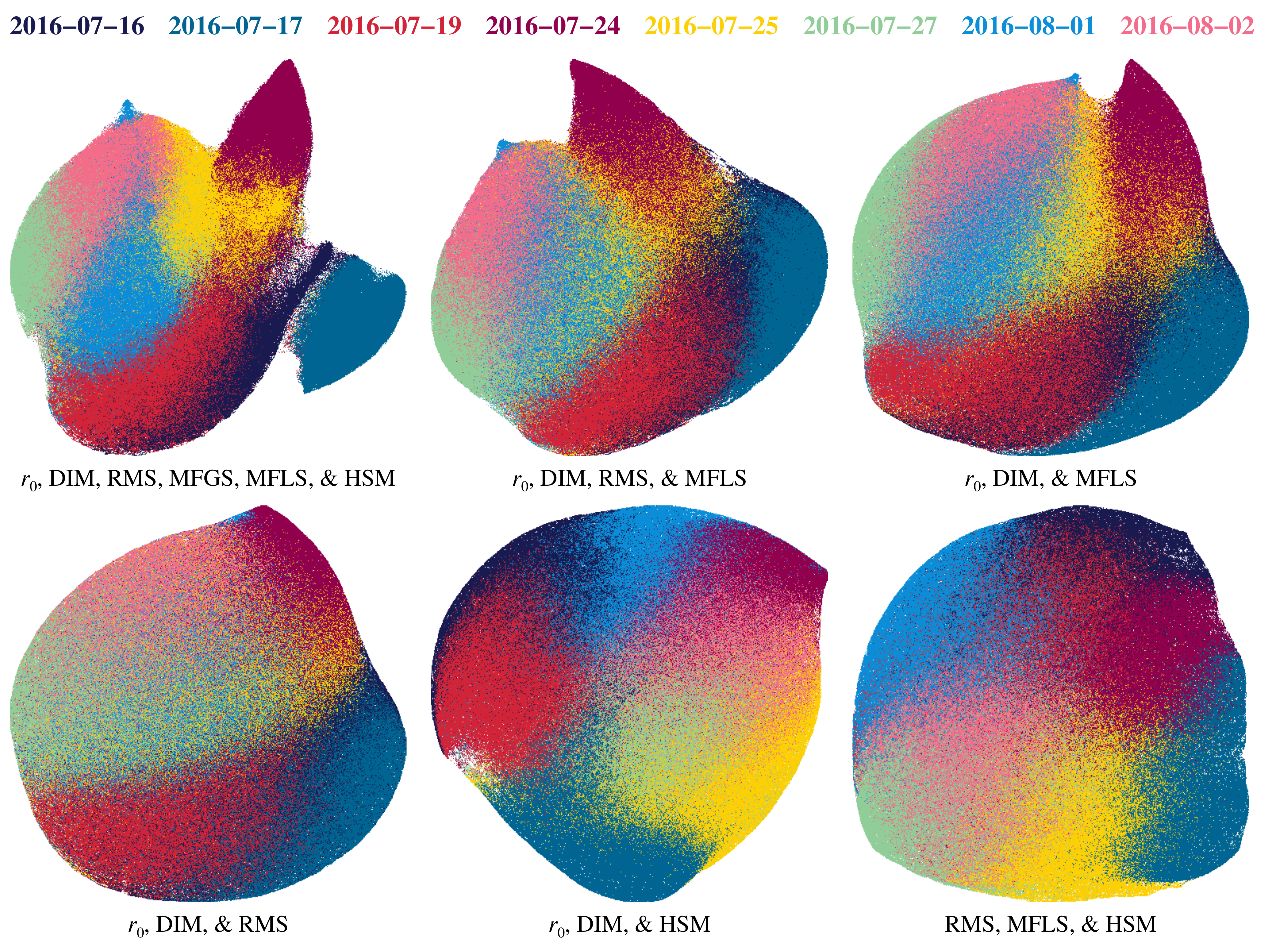}\smallskip
\caption{Two-dimensional UMAP projections for different combination seeing
    parameters and image quality metrics of isoplanatic patches for all eight observing days. The color code for the observing days is displayed at the top.}
\label{FIG12}
\end{figure}

The results for the G-band $\lambda$430.7~nm data depicted in the six panels of Figure~\ref{FIG12} were derived with the default settings, using an Euclidean metric and 15 neighbors to reduce the data down to two dimensions. The initial dimensions are given by the number of image quality metrics and seeing parameters. The two-dimensional UMAP projections are based on more than 1.5 million input features, \textit{i.e.}, various combinations of image quality metrics and seeing parameters, which are listed below each UMAP projection. The default settings already rendered adequate results. 

Using all six image quality metrics and seeing parameters, the observations on 2016 July~17 and~24 are distinctly separated from the other observing days, and they are tightly clustered at the periphery of the UMAP projection. This indicates that the seeing conditions might be very different on these two observing days. However, the other observing days are also located in well-defined regions of the UMAP projection though some overlap exists. Since they form an almost circular region, seeing conditions may have been very similar. Interestingly, the distance of an isoplanatic patch from the AO lock point does not enter into the projections, which supports the implicit assumption that seeing parameters and image quality metrics can be universally derived from three-dimensional spatio-temporal stacks of images with the size of the isoplanatic patch. Based on the poor performance of the Helmli-Scherer mean and on the fact that MFGS and MFLS have similar properties, the number of dimensions was reduced to four in the UMAP projection in the middle panel of the top row in 
Figure~\ref{FIG12}. The overall shape and morphology was preserved but the borders between different observing day become noisier. This trend becomes even more pronounced when moving from four to three dimensions in the remaining four plot panels. In particular, the seemingly preferable combination of Fried-parameter $r_0$, granular rms-contrast, and differential image motion performs poorly. Instead, at low dimensionality the Helmli-Scherer mean contributes to the separation of clusters. However, the preferred choice is the combination Fried-parameter $r_0$, differential image motion, and MFLS because the last metric performed better in the correlation analysis.

\begin{figure}[t]
\centering
\includegraphics[width=1.\textwidth]{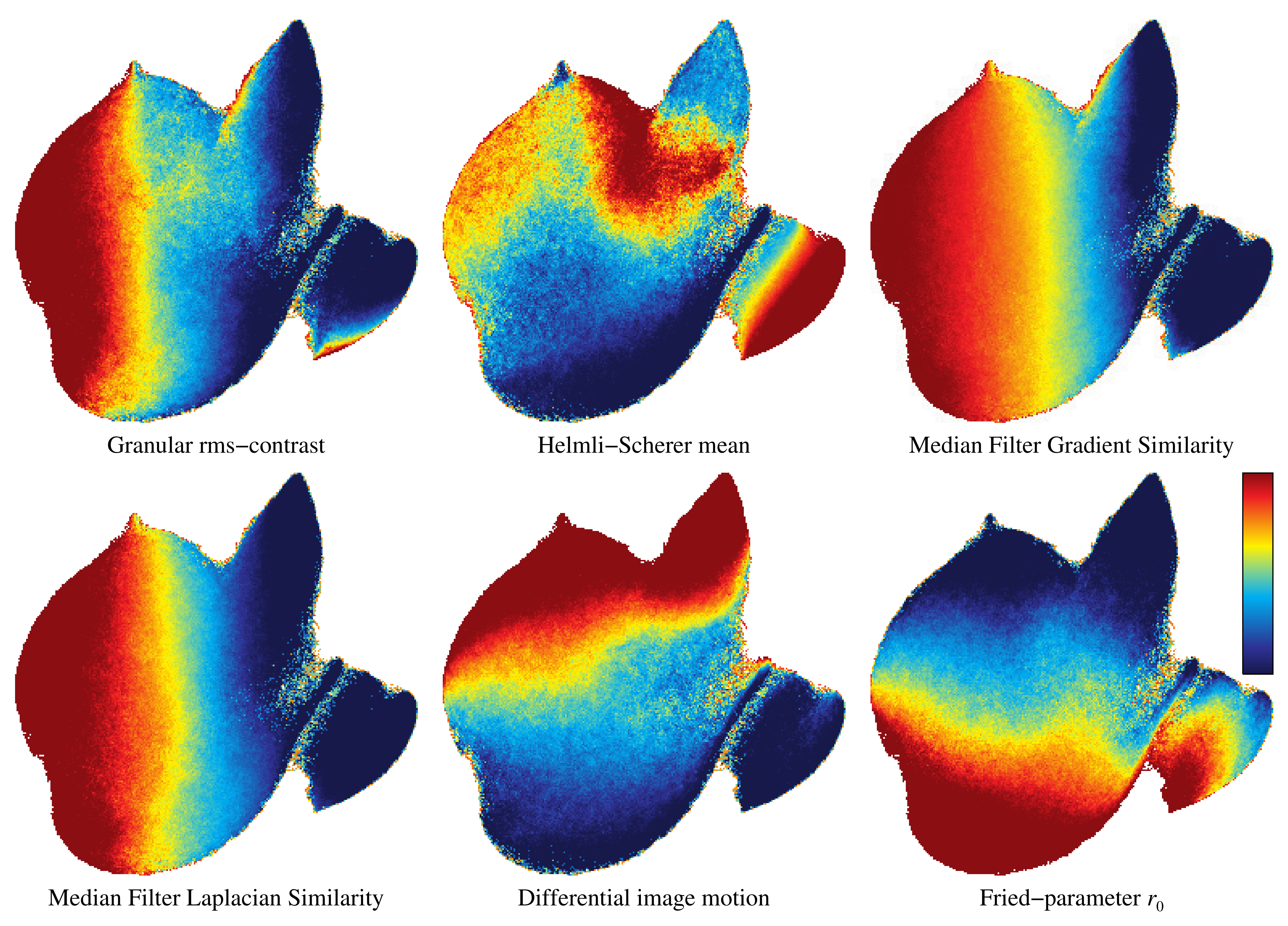}\smallskip
\caption{UMAP projections of all images quality metrics and seeing parameters. 
    The rainbow-colored scale bar in the lower-right panel refers, from blue to red, to the intervals RMS $\in$ [3.2\%, 5.2\%], HSM $\in$ [1.0049, 1.0057], MFGS $\in$ [0.935, 0.985], MFLS $\in$ [0.54, 0.62], DIM $\in$ [0.004~arcsec, 0.013~arcsec], and $r_0$ $\in$ [5.5~cm, 7.6~cm].}
\label{FIG13}
\end{figure}

The UMAP projections in Figure~\ref{FIG13} map all features (\textit{i.e.}, all images quality metrics and seeing parameters) to the two-dimensional projection space, illustrating how the dimensionality reduction is achieved. All six panels show distinct global gradients, which are horizontal for the granular rms-contrast, MFGS, and MFLS. Locally, the granular rms-contrast exhibits minor substructures for 2016 July~17 and~24. Close to vertical gradients are evident for Helmli-Scherer mean, differential image motion, and Fried-parameter $r_0$. However, the angle is about $45^\circ$ between the gradients for differential image motion, and Fried-parameter $r_0$, indicating that these two parameters plus one of the metrics with horizontal gradients are sufficient to describe the seeing variation at global scale. Only on 2016 July~17, Helmli-Scherer mean and Fried-parameter $r_0$ capture some distinctly different seeing contribution. Note the almost perpendicular gradients for this observing day. In addition, the UMAP projection of the Helmi-Scherer mean shows the most substructure, which warrants further scrutiny in future studies.

%
%

\section{Discussion}\label{SEC08}

The larger observed FOV of the HiFI imagers at the VTT, as compared to their usual location at the GREGOR solar telescope, made it possible to determine the field dependence of seeing parameters and image quality metrics covering a radial distance of 60~arcsec as measured from the lock point of the AO system. In July/August 2016, the VTT was operated without the entrance window, \textit{i.e.}, the vacuum vessel was not evacuated. Thermal or mechanical stresses in the entrance window are known to introduce static aberrations, which cannot be present in the observed images. Thus, the seeing comprises only contributions by the free atmosphere, the ground layer, the heated observing platform, and the slowly warming stagnant air inside the telescope tank. Telescope shake and guiding errors are additional causes of rigid image motion \citep{Tarbell1981}. A new entrance window was installed at the VTT in December 2019 but no additional measurements were possible so that the impact of the entrance window and of telescope seeing on AO performance cannot be judged. However, based on our experience with large-format CCD cameras at the VTT, even beyond a distance of 60~arcsec beyond the AO lock point, the AO system still aids image restoration techniques, if ever so slightly, by removing imperfections of the optical system and by at least partially correcting low-order modes.

While seeing improves with increasing wavelength, the image contrast decreases.
\citet{Scharmer2019} pointed out that increasing modes are needed for wavefront corrections when observing at shorter wavelengths. However, the maximum number of corrected modes is fixed for a given wavefront sensor and deformable mirror, and the central wavelength of the bandpass filter feeding light to the wavefront sensor is fixed and cannot be changed during observations adapting to the observed wavelength of the science instruments. Thus, the presented image quality metrics and seeing parameters include the superimposed wavelength-dependent effects of seeing variations and AO correction.

As pointed out by \citet{Grossmann-Doerth1969}, blurring, differential image motion, and rigid image motion occur simultaneously. However, their relative importance depend on the prevailing seeing conditions. Turbulent eddies larger than the telescope pupil result in image motion, whereas smaller blobs are responsible for blurring \citep{Brandt1969}. Temperature fluctuations and wind shear have a significant impact on the distribution of turbulent eddies, which increase in size with increasing altitude. Evaluating the atmospheric temperature structure $C_T$ \citep{Coulman1969} showed evidence that meso-scale thermal structures exist in the lower troposphere. While some of these show inhomogeneous temperature fine-structure that cause extreme non-uni\-form\-ity, others are almost free of it. Especially, near the heated ground convection process causes inhomogeneities in the temperature field. Typically, the height-dependent refractive index structure function $C^2_n(h)$ \citep{Tyson1998} is invoked to describe atmospheric turbulence caused by temperature fluctuations and wind motion. 

Evaluating $C^2_n (h)$ profiles, \citet{Masciadri2010} demonstrated that about 50\% of the seeing originates below $(80 \pm 15)$~m, and in total, 60\% is concentrated below 1~km. In addition, a secondary peak occurred at altitudes of about 10~km. This peak shows a seasonal dependence, with a shift to higher altitudes (14~km) and stronger turbulence in the winter. These results are in agreement with \citet{Martin1987} who associated the main contribution within the free atmosphere to a turbulence layer at 10\,--\,12~km, which is associated with the Jet stream. The range of 4\,--\,10~km possesses usually a uniform $C^2_n (h)$ profile, while up to an altitude of 4~km, the terrain can have severe influence on the atmospheric turbulence. Especially, telescopes located at mountain sides experience noteworthy shear motions, whereby the contribution of small and large eddies can be mixed up to several kilometers \citep{Scorer1963}. It is possible that turbulent layers are located in between laminar flows. A more recent investigation was carried out by \citet{Wang2018}, extending to altitudes of 16~km with a step size of 1~km. The results indicate three different locations where turbulence originates. In the lower troposphere 0\,--\,2~km, in the higher troposphere 3\,--\,6~km and in the upper most layer $\geq 7$~km with average percentages of 54.4, 39.8, 5.8\%, respectively. 

Linking features of the two-dimensional maps of the image quality metrics and seeing parameters to specific turbulence layers is a goal that is still beyond our grasp. Here, a much better statistic and simultaneous access to the wavefront sensor data is needed. The classification results of UMAP are promising and show a clear clustering of data points for each observing day. Thus, machine learning can be an essential tool to establish the aforementioned link. Establishing image quality criteria for data obtained with ground-based telescopes has impacts beyond solar observations. Recently, \citet{Kyono2020} applied convolutional neural networks (CNNs) to satellite images obtained at the \textit{Maui Space Surveillance Site} (MSSS). The CNN algorithm yields a quantitative score of image quality and predicts if image restoration will yield well-resolved images of satellites from the observed data.

%
%

\section{Conclusions}\label{SEC09}

Ground-based solar telescopes surpassed the 1-meter aperture limit of conventional evacuated telescopes. Real-time AO correction combined with \textit{post facto} image restoration techniques nowadays routinely delivers nearly diffraction-limited images. The primary objective of this study was to assess the seeing conditions at the VTT and the performance of its AO system. Descriptive seeing parameters and image quality metrics were analyzed with the practical goal of evaluating their potential use in data processing pipelines and data archives, and of providing guidance in selecting the most suitable image restoration technique for the prevailing seeing conditions. This small statistical study presented a detailed analysis of time- and field-dependent image quality metrics and seeing parameters, including rms-contrast of granulation, Helmli-Scherer mean, MFGS, MFLS, Fried-parameter $r_0$ derived with the spectral ratio technique, and differential image motion. Two-dimensional maps of these parameters exhibit a clear field dependence, decreasing with distance from the AO lock point for G-band and continuum images of the quiet Sun, which provides a uniform and isotropic target at disk center. However, contrast features such as sunspot, pores, plages, and bright points will leave an imprint in the two-dimensional maps, hampering a direct comparison. Geometric foreshortening close to the limb also makes a direct comparison more difficult. On the positive side, the results indicate a substantial AO correction beyond the AO lock point, which is much larger than the typical size of the isoplanatic patch, thus, significantly boosting image restoration. Once imaging systems, observational set-up, and observed scene on the Sun are characterized, suitable thresholds can be computed for choosing the appropriate image restoration technique, and databases can be objectively searched for the best datasets.


\begin{acks}
The Vacuum Tower telescope is operated by the German consortium of the Leibniz-Institut f\"ur Sonnenphysik (KIS) in Freiburg, the Leibniz-Institut f\"ur Astrophysik Potsdam (AIP), and the Max-Planck-Institut f\"ur Sonnensystemforschung (MPS) in G\"ottingen. This study was supported by grants VE~1112/1-1 and DE~787/5-1 of the Deutsche Forschungsgemeinschaft (DFG). In addition, the support by the European Commission's Horizon 2020 Program under grant agreements 824064 (ESCAPE -- European Science Cluster of Astronomy \& Particle physics ESFRI research infrastructures) and 824135 (SOLARNET -- Integrating High Resolution Solar Physics) is highly appreciated. This research has made use of NASA’s Astrophysics Data System. The authors would like to thank Dr.\ Michael Weber for helpful for his comments, suggestions, and careful reading of the manuscript.
\medskip

\noindent\textbf{Disclosure of Potential Conflicts of Interest}$\quad$ The authors declare that they have no conflicts of interest. 
\end{acks}

%
%

\end{article}


\begin{thebibliography}{69}
\ifx\bisbn     \undefined \def\bisbn  #1{ISBN #1}\fi
\ifx\binits    \undefined \def\binits#1{#1}\fi
\ifx\bauthor   \undefined \def\bauthor#1{#1}\fi
\ifx\batitle   \undefined \def\batitle#1{#1}\fi
\ifx\bjtitle   \undefined \def\bjtitle#1{\textit{#1}}\fi
\ifx\bvolume   \undefined \def\bvolume#1{\textbf{#1}}\fi
\ifx\byear     \undefined \def\byear#1{#1}\fi
\ifx\bissue    \undefined \def\bissue#1{#1}\fi
\ifx\bfpage    \undefined \def\bfpage#1{#1}\fi
\ifx\blpage    \undefined \def\blpage #1{#1}\fi
\ifx\burl      \undefined \def\burl#1{\textsf{#1}}\fi
\ifx\href      \undefined \def\href#1#2{\textsf{#2}}\fi
\ifx\betal     \undefined \def\betal{\textit{et al.}}\fi
\ifx\bctitle   \undefined \def\bctitle#1{#1}\fi
\ifx\beditor   \undefined \def\beditor#1{#1}\fi
\ifx\bbtitle   \undefined \def\bbtitle#1{\textit{#1}}\fi
\ifx\bedition  \undefined \def\bedition#1{#1}\fi
\ifx\bseriesno \undefined \def\bseriesno#1{\textbf{#1}}\fi
\ifx\blocation \undefined \def\blocation#1{#1}\fi
\ifx\bsertitle \undefined \def\bsertitle#1{\textit{#1}}\fi
\ifx\bsnm      \undefined \def\bsnm#1{#1}\fi
\ifx\bsuffix   \undefined \def\bsuffix#1{#1}\fi
\ifx\bparticle \undefined \def\bparticle#1{#1}\fi
\ifx\barticle  \undefined \def\barticle#1{}\fi
\ifx\binstitute  \undefined \def\binstitute#1{#1}\fi
\ifx\bpublisher  \undefined \def\bpublisher#1{#1}\fi
\ifx\doiurl    \undefined
  \def\doiurl#1{\href{http://dx.doi.org/#1}{\textsf{DOI}}}\fi
\ifx\arxivurl  \undefined
  \def\arxivurl#1{\href{http://arxiv.org/abs/#1}{\textsf{arXiv}}}\fi
\ifx\adsurl    \undefined
  \def\adsurl#1{\href{http://adsabs.harvard.edu/abs/#1}{\textsf{ADS}}}\fi
\ifx\botherref \undefined \def\botherref#1{}\fi
\ifx\url       \undefined \def\url#1{\textsf{#1}}\fi
\ifx\bchapter  \undefined \def\bchapter#1{}\fi
\ifx\bbook     \undefined \def\bbook#1{}\fi
\ifx\bcomment  \undefined \def\bcomment#1{#1}\fi
\ifx\oauthor   \undefined \def\oauthor#1{#1}\fi
\ifx\citeauthoryear \undefined\def \citeauthoryear#1{#1}\fi
\ifx\endbibitem\undefined \def\endbibitem{}\fi
\ifx\bconflocation  \undefined \def\bconflocation#1{#1} \fi

\bibitem[\protect\citeauthoryear{{Beckers}}{2001}]{Beckers2001}
\begin{barticle}
\bauthor{\bsnm{{Beckers}}, \binits{J.M.}}:
\byear{2001},
\batitle{{A Seeing Monitor for Solar and Other Extended Object Observations}}.
\bjtitle{Exp. Astron.}
\bvolume{12},
\bfpage{1}.
\end{barticle}
\endbibitem

\bibitem[\protect\citeauthoryear{{Berkefeld}
  \textit{et~al.}}{2010}]{Berkefeld2010}
\begin{barticle}
\bauthor{\bsnm{{Berkefeld}}, \binits{T.}},
\bauthor{\bsnm{{Soltau}}, \binits{D.}},
\bauthor{\bsnm{{Schmidt}}, \binits{D.}},
\bauthor{\bsnm{{von der L{\"u}he}}, \binits{O.}}:
\byear{2010},
\batitle{{Adaptive Optics Development at the German Solar Telescopes}}.
\bjtitle{Appl. Opt.}
\bvolume{49},
\bfpage{G155}.
\doiurl{10.1364/AO.49.00G155}.
\end{barticle}
\endbibitem

\bibitem[\protect\citeauthoryear{{Brandt}}{1969}]{Brandt1969}
\begin{barticle}
\bauthor{\bsnm{{Brandt}}, \binits{P.N.}}:
\byear{1969},
\batitle{{Frequency Spectra of Solar Image Motion}}.
\bjtitle{Solar Phys.}
\bvolume{7},
\bfpage{187}.
\doiurl{10.1007/BF00224897}.
\end{barticle}
\endbibitem

\bibitem[\protect\citeauthoryear{{Brandt} and {W{\"o}hl}}{1982}]{Brandt1982}
\begin{barticle}
\bauthor{\bsnm{{Brandt}}, \binits{P.N.}},
\bauthor{\bsnm{{W{\"o}hl}}, \binits{H.}}:
\byear{1982},
\batitle{{Solar Site-testing Campaign of JOSO on the Canary Islands in 1979}}.
\bjtitle{Astron. Astrophys.}
\bvolume{109},
\bfpage{77}.
\end{barticle}
\endbibitem

\bibitem[\protect\citeauthoryear{{Cagigal} and {Canales}}{2000}]{Cagigal2000}
\begin{barticle}
\bauthor{\bsnm{{Cagigal}}, \binits{M.P.}},
\bauthor{\bsnm{{Canales}}, \binits{V.E.}}:
\byear{2000},
\batitle{{Generalized Fried Parameter after Adaptive Optics Partial Wave-front
  Compensation}}.
\bjtitle{J. Opt. Soc. Am. A}
\bvolume{17},
\bfpage{903}.
\doiurl{10.1364/JOSAA.17.000903}.
\end{barticle}
\endbibitem

\bibitem[\protect\citeauthoryear{{Coulman}}{1969}]{Coulman1969}
\begin{barticle}
\bauthor{\bsnm{{Coulman}}, \binits{C.E.}}:
\byear{1969},
\batitle{{A Quantitative Treatment of Solar `Seeing'}}.
\bjtitle{Solar Phys.}
\bvolume{7},
\bfpage{122}.
\doiurl{10.1007/BF00148409}.
\end{barticle}
\endbibitem

\bibitem[\protect\citeauthoryear{{Danilovic}
  \textit{et~al.}}{2008}]{Danilovic2008}
\begin{barticle}
\bauthor{\bsnm{{Danilovic}}, \binits{S.}},
\bauthor{\bsnm{{Gandorfer}}, \binits{A.}},
\bauthor{\bsnm{{Lagg}}, \binits{A.}},
\bauthor{\bsnm{{Sch{\"u}ssler}}, \binits{M.}},
\bauthor{\bsnm{{Solanki}}, \binits{S.K.}},
\bauthor{\bsnm{{V{\"o}gler}}, \binits{A.}},
\bauthor{\bsnm{{Katsukawa}}, \binits{Y.}},
\bauthor{\bsnm{{Tsuneta}}, \binits{S.}}:
\byear{2008},
\batitle{{The Intensity Contrast of Solar Granulation: Comparing Hinode SP
  Results with MHD Simulations}}.
\bjtitle{Astron. Astrophys.}
\bvolume{484},
\bfpage{L17}.
\doiurl{10.1051/0004-6361:200809857}.
\end{barticle}
\endbibitem

\bibitem[\protect\citeauthoryear{{Deng} \textit{et~al.}}{2015}]{Deng2015}
\begin{barticle}
\bauthor{\bsnm{{Deng}}, \binits{H.}},
\bauthor{\bsnm{{Zhang}}, \binits{D.}},
\bauthor{\bsnm{{Wang}}, \binits{T.}},
\bauthor{\bsnm{{Ji}}, \binits{K.}},
\bauthor{\bsnm{{Wang}}, \binits{F.}},
\bauthor{\bsnm{{Liu}}, \binits{Z.}},
\bauthor{\bsnm{{Xiang}}, \binits{Y.}},
\bauthor{\bsnm{{Jin}}, \binits{Z.}},
\bauthor{\bsnm{{Cao}}, \binits{W.}}:
\byear{2015},
\batitle{{Objective Image-quality Assessment for High-resolution Photospheric
  Images by Median Filter Gradient Similarity}}.
\bjtitle{Solar Phys.}
\bvolume{290},
\bfpage{1479}.
\doiurl{10.1007/s11207-015-0676-1}.
\end{barticle}
\endbibitem

\bibitem[\protect\citeauthoryear{{Denker}, {Tritschler}, and
  {L{\"o}fdahl}}{2015}]{Denker2015}
\begin{bchapter}
\bauthor{\bsnm{{Denker}}, \binits{C.}},
\bauthor{\bsnm{{Tritschler}}, \binits{A.}},
\bauthor{\bsnm{{L{\"o}fdahl}}, \binits{M.}}:
\byear{2015},
\bctitle{{Image Restoration}}.
In: \beditor{\bsnm{{Hoffman}}, \binits{C.}},
\beditor{\bsnm{{Driggers}}, \binits{R.}} (eds.)
\bbtitle{{Encyclopedia of Optical and Photonic Engineering}},
\bedition{{2$^\mathrm{nd}$}} edn.,
\bpublisher{{CRC Press}},
\blocation{{Boca Raton, Florida}},
\bfpage{1}.
\doiurl{doi:10.1081/E-EOE2-120038539}.
\end{bchapter}
\endbibitem

\bibitem[\protect\citeauthoryear{{Denker} \textit{et~al.}}{2005}]{Denker2005b}
\begin{barticle}
\bauthor{\bsnm{{Denker}}, \binits{C.}},
\bauthor{\bsnm{{Mascarinas}}, \binits{D.}},
\bauthor{\bsnm{{Xu}}, \binits{Y.}},
\bauthor{\bsnm{{Cao}}, \binits{W.}},
\bauthor{\bsnm{{Yang}}, \binits{G.}},
\bauthor{\bsnm{{Wang}}, \binits{H.}},
\bauthor{\bsnm{{Goode}}, \binits{P.R.}},
\bauthor{\bsnm{{Rimmele}}, \binits{T.R.}}:
\byear{2005},
\batitle{{High-spatial Resolution Imaging Combining High-order Adaptive Optics,
  Frame Selection, and Speckle Masking Reconstruction}}.
\bjtitle{Solar Phys.}
\bvolume{227},
\bfpage{217}.
\doiurl{10.1007/s11207-005-1108-4}.
\end{barticle}
\endbibitem

\bibitem[\protect\citeauthoryear{{Denker} \textit{et~al.}}{2007a}]{Denker2007b}
\begin{barticle}
\bauthor{\bsnm{{Denker}}, \binits{C.}},
\bauthor{\bsnm{{Tritschler}}, \binits{A.}},
\bauthor{\bsnm{{Rimmele}}, \binits{T.R.}},
\bauthor{\bsnm{{Richards}}, \binits{K.}},
\bauthor{\bsnm{{Hegwer}}, \binits{S.L.}},
\bauthor{\bsnm{{W{\"o}ger}}, \binits{F.}}:
\byear{2007}a,
\batitle{{Adaptive Optics at the Big Bear Solar Observatory: Instrument
  Description and First Observations}}.
\bjtitle{Publ. Astron. Soc. Pac.}
\bvolume{119},
\bfpage{170}.
\doiurl{10.1086/512493}.
\end{barticle}
\endbibitem

\bibitem[\protect\citeauthoryear{{Denker} \textit{et~al.}}{2007b}]{Denker2007a}
\begin{barticle}
\bauthor{\bsnm{{Denker}}, \binits{C.}},
\bauthor{\bsnm{{Deng}}, \binits{N.}},
\bauthor{\bsnm{{Rimmele}}, \binits{T.R.}},
\bauthor{\bsnm{{Tritschler}}, \binits{A.}},
\bauthor{\bsnm{{Verdoni}}, \binits{A.}}:
\byear{2007}b,
\batitle{{Field-dependent Adaptive Optics Correction Derived with the Spectral
  Ratio Technique}}.
\bjtitle{Solar Phys.}
\bvolume{241},
\bfpage{411}.
\doiurl{10.1007/s11207-007-0315-6}.
\end{barticle}
\endbibitem

\bibitem[\protect\citeauthoryear{{Denker} \textit{et~al.}}{2012}]{Denker2012}
\begin{barticle}
\bauthor{\bsnm{{Denker}}, \binits{C.}},
\bauthor{\bsnm{{von der L{\"u}he}}, \binits{O.}},
\bauthor{\bsnm{{Feller}}, \binits{A.}},
\bauthor{\bsnm{{Arlt}}, \binits{K.}},
\bauthor{\bsnm{{Balthasar}}, \binits{H.}},
\bauthor{\bsnm{{Bauer}}, \binits{S.-M.}},
\bauthor{\bsnm{{Bello Gonz{\'a}lez}}, \binits{N.}},
\bauthor{\bsnm{{Berkefeld}}, \binits{T.}},
\bauthor{\bsnm{{Caligari}}, \binits{P.}},
\bauthor{\bsnm{{Collados}}, \binits{M.}},
\bauthor{\bsnm{{Fischer}}, \binits{A.}},
\bauthor{\bsnm{{Granzer}}, \binits{T.}},
\bauthor{\bsnm{{Hahn}}, \binits{T.}},
\bauthor{\bsnm{{Halbgewachs}}, \binits{C.}},
\bauthor{\bsnm{{Heidecke}}, \binits{F.}},
\bauthor{\bsnm{{Hofmann}}, \binits{A.}},
\bauthor{\bsnm{{Kentischer}}, \binits{T.}},
\bauthor{\bsnm{{Klva{\v n}a}}, \binits{M.}},
\bauthor{\bsnm{{Kneer}}, \binits{F.}},
\bauthor{\bsnm{{Lagg}}, \binits{A.}},
\bauthor{\bsnm{{Nicklas}}, \binits{H.}},
\bauthor{\bsnm{{Popow}}, \binits{E.}},
\bauthor{\bsnm{{Puschmann}}, \binits{K.G.}},
\bauthor{\bsnm{{Rendtel}}, \binits{J.}},
\bauthor{\bsnm{{Schmidt}}, \binits{D.}},
\bauthor{\bsnm{{Schmidt}}, \binits{W.}},
\bauthor{\bsnm{{Sobotka}}, \binits{M.}},
\bauthor{\bsnm{{Solanki}}, \binits{S.K.}},
\bauthor{\bsnm{{Soltau}}, \binits{D.}},
\bauthor{\bsnm{{Staude}}, \binits{J.}},
\bauthor{\bsnm{{Strassmeier}}, \binits{K.G.}},
\bauthor{\bsnm{{Volkmer}}, \binits{R.}},
\bauthor{\bsnm{{Waldmann}}, \binits{T.}},
\bauthor{\bsnm{{Wiehr}}, \binits{E.}},
\bauthor{\bsnm{{Wittmann}}, \binits{A.D.}},
\bauthor{\bsnm{{Woche}}, \binits{M.}}:
\byear{2012},
\batitle{{A Retrospective of the GREGOR Solar Telescope in Scientific
  Literature}}.
\bjtitle{Astron. Nachr.}
\bvolume{333},
\bfpage{810}.
\doiurl{10.1002/asna.201211728}.
\end{barticle}
\endbibitem

\bibitem[\protect\citeauthoryear{{Denker} \textit{et~al.}}{2018a}]{Denker2018a}
\begin{barticle}
\bauthor{\bsnm{{Denker}}, \binits{C.}},
\bauthor{\bsnm{{Kuckein}}, \binits{C.}},
\bauthor{\bsnm{{Verma}}, \binits{M.}},
\bauthor{\bsnm{{Gonz{\'a}lez Manrique}}, \binits{S.J.}},
\bauthor{\bsnm{{Diercke}}, \binits{A.}},
\bauthor{\bsnm{{Enke}}, \binits{H.}},
\bauthor{\bsnm{{Klar}}, \binits{J.}},
\bauthor{\bsnm{{Balthasar}}, \binits{H.}},
\bauthor{\bsnm{{Louis}}, \binits{R.E.}},
\bauthor{\bsnm{{Dineva}}, \binits{E.}}:
\byear{2018}a,
\batitle{{Data Analysis and Management for High-resolution Solar Physics --
  Image Restoration and Imaging Spectroscopy at the GREGOR Solar Telescope}}.
\bjtitle{Astrophys. J. Suppl. Ser.}
\bvolume{236},
\bfpage{5}.
\doiurl{10.3847/1538-4365/aab773}.
\end{barticle}
\endbibitem

\bibitem[\protect\citeauthoryear{{Denker} \textit{et~al.}}{2018b}]{Denker2018b}
\begin{barticle}
\bauthor{\bsnm{{Denker}}, \binits{C.}},
\bauthor{\bsnm{{Dineva}}, \binits{E.}},
\bauthor{\bsnm{{Balthasar}}, \binits{H.}},
\bauthor{\bsnm{{Verma}}, \binits{M.}},
\bauthor{\bsnm{{Kuckein}}, \binits{C.}},
\bauthor{\bsnm{{Diercke}}, \binits{A.}},
\bauthor{\bsnm{{Gonz{\'a}lez Manrique}}, \binits{S.J.}}:
\byear{2018}b,
\batitle{{Image Quality in High-resolution and High-cadence Solar Imaging}}.
\bjtitle{Solar Phys.}
\bvolume{293},
\bfpage{44}.
\doiurl{10.1007/s11207-018-1261-1}.
\end{barticle}
\endbibitem

\bibitem[\protect\citeauthoryear{{Fried}}{1966}]{Fried1966}
\begin{barticle}
\bauthor{\bsnm{{Fried}}, \binits{D.L.}}:
\byear{1966},
\batitle{{Optical Resolution Through a Randomly Inhomogeneous Medium for Very
  Long and Very Short Exposures}}.
\bjtitle{J. Opt. Soc. Am. A}
\bvolume{56},
\bfpage{1372}.
\doiurl{10.1364/JOSA.56.001372}.
\end{barticle}
\endbibitem

\bibitem[\protect\citeauthoryear{{Fried} and {Mevers}}{1974}]{Fried1974}
\begin{barticle}
\bauthor{\bsnm{{Fried}}, \binits{D.L.}},
\bauthor{\bsnm{{Mevers}}, \binits{G.E.}}:
\byear{1974},
\batitle{{Evaluation of $r_{0}$ for Propagation Down through the Atmosphere}}.
\bjtitle{Appl. Opt.}
\bvolume{13},
\bfpage{2620}.
\doiurl{10.1364/AO.13.002620}.
\end{barticle}
\endbibitem

\bibitem[\protect\citeauthoryear{{Gallagher}, {Moon}, and
  {Wang}}{2002}]{Gallagher2002a}
\begin{barticle}
\bauthor{\bsnm{{Gallagher}}, \binits{P.T.}},
\bauthor{\bsnm{{Moon}}, \binits{Y.-J.}},
\bauthor{\bsnm{{Wang}}, \binits{H.}}:
\byear{2002},
\batitle{{Active-region Monitoring and Flare Forecasting I. Data Processing and
  First Results}}.
\bjtitle{Solar Phys.}
\bvolume{209},
\bfpage{171}.
\end{barticle}
\endbibitem

\bibitem[\protect\citeauthoryear{{Grossmann-Doerth}}{1969}]{Grossmann-Doerth1969}
\begin{barticle}
\bauthor{\bsnm{{Grossmann-Doerth}}, \binits{U.}}:
\byear{1969},
\batitle{{On Astronomical Seeing: The Single Schlieren Model}}.
\bjtitle{Solar Phys.}
\bvolume{9},
\bfpage{210}.
\doiurl{10.1007/BF00145743}.
\end{barticle}
\endbibitem

\bibitem[\protect\citeauthoryear{{Helmli} and {Scherer}}{2001}]{Helmli2001}
\begin{bchapter}
\bauthor{\bsnm{{Helmli}}, \binits{F.S.}},
\bauthor{\bsnm{{Scherer}}, \binits{S.}}:
\byear{2001},
\bctitle{{Adaptive Shape from Focus with an Error Estimation in Light
  Microscopy}}.
In: \beditor{\bsnm{{Lon\v{c}ari\'c}}, \binits{S.}},
\beditor{\bsnm{{Babi\'c}}, \binits{H.}} (eds.)
\bbtitle{Proceedings of the 2nd International Symposium on Image and Signal
  Processing and Analysis. In Conjunction with 23rd International Conference on
  Information Technology Interfaces},
\bsertitle{IEEE Cat. No.01EX480},
\bfpage{188}.
\doiurl{10.1109/ISPA.2001.938626}.
\end{bchapter}
\endbibitem

\bibitem[\protect\citeauthoryear{{Hill} \textit{et~al.}}{2006}]{Hill2006}
\begin{bchapter}
\bauthor{\bsnm{{Hill}}, \binits{F.}},
\bauthor{\bsnm{{Beckers}}, \binits{J.}},
\bauthor{\bsnm{{Brandt}}, \binits{P.}},
\bauthor{\bsnm{{Briggs}}, \binits{J.}},
\bauthor{\bsnm{{Brown}}, \binits{T.}},
\bauthor{\bsnm{{Brown}}, \binits{W.}},
\bauthor{\bsnm{{Collados}}, \binits{M.}},
\bauthor{\bsnm{{Denker}}, \binits{C.}},
\bauthor{\bsnm{{Fletcher}}, \binits{S.}},
\bauthor{\bsnm{{Hegwer}}, \binits{S.}},
\bauthor{\bsnm{{Horst}}, \binits{T.}},
\bauthor{\bsnm{{Komsa}}, \binits{M.}},
\bauthor{\bsnm{{Kuhn}}, \binits{J.}},
\bauthor{\bsnm{{Lecinski}}, \binits{A.}},
\bauthor{\bsnm{{Lin}}, \binits{H.}},
\bauthor{\bsnm{{Oncley}}, \binits{S.}},
\bauthor{\bsnm{{Penn}}, \binits{M.}},
\bauthor{\bsnm{{Radick}}, \binits{R.}},
\bauthor{\bsnm{{Rimmele}}, \binits{T.}},
\bauthor{\bsnm{{Socas-Navarro}}, \binits{H.}},
\bauthor{\bsnm{{Streander}}, \binits{K.}}:
\byear{2006},
\bctitle{{Site Testing for the Advanced Technology Solar Telescope}}.
In: \beditor{\bsnm{{Stepp}}, \binits{L.M.}} (ed.)
\bbtitle{Ground-based and Airborne Telescopes},
\bsertitle{Proc. SPIE}
\bseriesno{6267},
\bfpage{62671T}.
\end{bchapter}
\endbibitem

\bibitem[\protect\citeauthoryear{{Hunt}, {Fright}, and
  {Bates}}{1983}]{Hunt1983}
\begin{barticle}
\bauthor{\bsnm{{Hunt}}, \binits{B.R.}},
\bauthor{\bsnm{{Fright}}, \binits{W.R.}},
\bauthor{\bsnm{{Bates}}, \binits{R.H.T.}}:
\byear{1983},
\batitle{{Analysis of the Shift-and-add Method for Imaging through Turbulent
  Media}}.
\bjtitle{J. Opt. Soc. Am.}
\bvolume{73},
\bfpage{456}.
\doiurl{10.1364/JOSA.73.000456}.
\end{barticle}
\endbibitem

\bibitem[\protect\citeauthoryear{{Korff}}{1973}]{Korff1973}
\begin{barticle}
\bauthor{\bsnm{{Korff}}, \binits{D.}}:
\byear{1973},
\batitle{{Analysis of a Method for Obtaining Near-diffraction-limited
  Information in the Presence of Atmospheric Turbulence}}.
\bjtitle{J. Opt. Soc. Am. A}
\bvolume{63},
\bfpage{971}.
\end{barticle}
\endbibitem

\bibitem[\protect\citeauthoryear{{Kosugi} \textit{et~al.}}{2007}]{Kosugi2007}
\begin{barticle}
\bauthor{\bsnm{{Kosugi}}, \binits{T.}},
\bauthor{\bsnm{{Matsuzaki}}, \binits{K.}},
\bauthor{\bsnm{{Sakao}}, \binits{T.}},
\bauthor{\bsnm{{Shimizu}}, \binits{T.}},
\bauthor{\bsnm{{Sone}}, \binits{Y.}},
\bauthor{\bsnm{{Tachikawa}}, \binits{S.}},
\bauthor{\bsnm{{Hashimoto}}, \binits{T.}},
\bauthor{\bsnm{{Minesugi}}, \binits{K.}},
\bauthor{\bsnm{{Ohnishi}}, \binits{A.}},
\bauthor{\bsnm{{Yamada}}, \binits{T.}},
\bauthor{\bsnm{{Tsuneta}}, \binits{S.}},
\bauthor{\bsnm{{Hara}}, \binits{H.}},
\bauthor{\bsnm{{Ichimoto}}, \binits{K.}},
\bauthor{\bsnm{{Suematsu}}, \binits{Y.}},
\bauthor{\bsnm{{Shimojo}}, \binits{M.}},
\bauthor{\bsnm{{Watanabe}}, \binits{T.}},
\bauthor{\bsnm{{Shimada}}, \binits{S.}},
\bauthor{\bsnm{{Davis}}, \binits{J.M.}},
\bauthor{\bsnm{{Hill}}, \binits{L.D.}},
\bauthor{\bsnm{{Owens}}, \binits{J.K.}},
\bauthor{\bsnm{{Title}}, \binits{A.M.}},
\bauthor{\bsnm{{Culhane}}, \binits{J.L.}},
\bauthor{\bsnm{{Harra}}, \binits{L.K.}},
\bauthor{\bsnm{{Doschek}}, \binits{G.A.}},
\bauthor{\bsnm{{Golub}}, \binits{L.}}:
\byear{2007},
\batitle{{The Hinode (Solar-B) Mission: An Overview}}.
\bjtitle{Solar Phys.}
\bvolume{243},
\bfpage{3}.
\doiurl{10.1007/s11207-007-9014-6}.
\end{barticle}
\endbibitem

\bibitem[\protect\citeauthoryear{{Kuckein}
  \textit{et~al.}}{2017}]{Kuckein2017a}
\begin{bchapter}
\bauthor{\bsnm{{Kuckein}}, \binits{C.}},
\bauthor{\bsnm{{Denker}}, \binits{C.}},
\bauthor{\bsnm{{Verma}}, \binits{M.}},
\bauthor{\bsnm{{Balthasar}}, \binits{H.}},
\bauthor{\bsnm{{Gonz{\'a}lez Manrique}}, \binits{S.J.}},
\bauthor{\bsnm{{Louis}}, \binits{R.E.}},
\bauthor{\bsnm{{Diercke}}, \binits{A.}}:
\byear{2017},
\bctitle{{sTools -- A Data Reduction Pipeline for the GREGOR Fabry-P{\'e}rot
  Interferometer and the High-resolution Fast Imager at the GREGOR Solar
  Telescope}}.
In: \beditor{\bsnm{{Vargas Dom{\'\i}nguez}}, \binits{S.}},
\beditor{\bsnm{{Kosovichev}}, \binits{A.G.}},
\beditor{\bsnm{{Harra}}, \binits{L.}},
\beditor{\bsnm{{Antolin}}, \binits{P.}} (eds.)
\bbtitle{Fine Structure and Dynamics of the Solar Atmosphere},
\bsertitle{IAU Symp.}
\bseriesno{327},
\bfpage{20}.
\doiurl{10.1017/S1743921317000114}.
\end{bchapter}
\endbibitem

\bibitem[\protect\citeauthoryear{{Kyono} \textit{et~al.}}{2020}]{Kyono2020}
\begin{barticle}
\bauthor{\bsnm{{Kyono}}, \binits{T.}},
\bauthor{\bsnm{{Lucas}}, \binits{J.}},
\bauthor{\bsnm{{Werth}}, \binits{M.}},
\bauthor{\bsnm{{Calef}}, \binits{B.}},
\bauthor{\bsnm{{McQuaid}}, \binits{I.}},
\bauthor{\bsnm{{Fletcher}}, \binits{J.}}:
\byear{2020},
\batitle{{Machine Learning for Quality Assessment of Ground-based Optical
  Images of Satellites}}.
\bjtitle{Opt. Eng.}
\bvolume{59},
\bfpage{1}.
\doiurl{10.1117/1.OE.59.5.051403}.
\end{barticle}
\endbibitem

\bibitem[\protect\citeauthoryear{{L{\"o}fdahl}}{2016}]{Loefdahl2016}
\begin{bchapter}
\bauthor{\bsnm{{L{\"o}fdahl}}, \binits{M.}}:
\byear{2016},
\bctitle{{A Comparison of Solar Image Restoration Techniques for SST/CRISP
  Data}}.
In: \beditor{\bsnm{{Dorotovic}}, \binits{I.}},
\beditor{\bsnm{{Fischer}}, \binits{C.E.}},
\beditor{\bsnm{{Temmer}}, \binits{M.}} (eds.)
\bbtitle{Coimbra Solar Physics Meeting: Ground-based Solar Observations in the
  Space Instrumentation Era},
\bsertitle{ASP Conf. Ser.}
\bseriesno{504},
\bfpage{111}.
\end{bchapter}
\endbibitem

\bibitem[\protect\citeauthoryear{{L{\"o}fdahl}}{2002}]{Loefdahl2002}
\begin{bchapter}
\bauthor{\bsnm{{L{\"o}fdahl}}, \binits{M.G.}}:
\byear{2002},
\bctitle{{Multi-frame Blind Deconvolution with Linear Equality Constraints}}.
In: \beditor{\bsnm{{Bones}}, \binits{P.J.}},
\beditor{\bsnm{{Fiddy}}, \binits{M.A.}},
\beditor{\bsnm{{Millane}}, \binits{R.P.}} (eds.)
\bbtitle{Image Reconstruction from Incomplete Data},
\bsertitle{Proc. SPIE}
\bseriesno{4792},
\bfpage{146}.
\doiurl{10.1117/12.451791}.
\end{bchapter}
\endbibitem

\bibitem[\protect\citeauthoryear{{L{\"o}fdahl}}{2010}]{Loefdahl2010}
\begin{barticle}
\bauthor{\bsnm{{L{\"o}fdahl}}, \binits{M.G.}}:
\byear{2010},
\batitle{{Evaluation of Image-shift Measurement Algorithms for Solar
  Shack-Hartmann Wavefront Sensors}}.
\bjtitle{Astron. Astrophys.}
\bvolume{524},
\bfpage{A90}.
\doiurl{10.1051/0004-6361/201015331}.
\end{barticle}
\endbibitem

\bibitem[\protect\citeauthoryear{{L{\"o}fdahl}, {van Noort}, and
  {Denker}}{2007}]{Loefdahl2007}
\begin{bchapter}
\bauthor{\bsnm{{L{\"o}fdahl}}, \binits{M.G.}},
\bauthor{\bsnm{{van Noort}}, \binits{M.J.}},
\bauthor{\bsnm{{Denker}}, \binits{C.}}:
\byear{2007},
\bctitle{{Solar Image Restoration}}.
In: \beditor{\bsnm{{Kneer}}, \binits{F.}},
\beditor{\bsnm{{Puschmann}}, \binits{K.G.}},
\beditor{\bsnm{{Wittmann}}, \binits{A.D.}} (eds.)
\bbtitle{Modern Solar Facilities -- Advanced Solar Science},
\bfpage{119}.
\end{bchapter}
\endbibitem

\bibitem[\protect\citeauthoryear{{Markwardt}}{2009}]{Markwardt2009}
\begin{bchapter}
\bauthor{\bsnm{{Markwardt}}, \binits{C.B.}}:
\byear{2009},
\bctitle{{Non-linear Least-squares Fitting in IDL with MPFIT}}.
In: \beditor{\bsnm{{Bohlender}}, \binits{D.A.}},
\beditor{\bsnm{{Durand}}, \binits{D.}},
\beditor{\bsnm{{Dowler}}, \binits{P.}} (eds.)
\bbtitle{Astronomical Data Analysis Software and Systems XVIII},
\bsertitle{ASP Conf. Ser.}
\bseriesno{411},
\bfpage{251}.
\end{bchapter}
\endbibitem

\bibitem[\protect\citeauthoryear{{Martin}}{1987}]{Martin1987}
\begin{barticle}
\bauthor{\bsnm{{Martin}}, \binits{H.M.}}:
\byear{1987},
\batitle{{Image Motion as a Measure of Seeing Quality}}.
\bjtitle{Publ. Astron. Soc. Pac.}
\bvolume{99},
\bfpage{1360}.
\doiurl{10.1086/132126}.
\end{barticle}
\endbibitem

\bibitem[\protect\citeauthoryear{{Masciadri}
  \textit{et~al.}}{2010}]{Masciadri2010}
\begin{bchapter}
\bauthor{\bsnm{{Masciadri}}, \binits{E.}},
\bauthor{\bsnm{{Stoesz}}, \binits{J.}},
\bauthor{\bsnm{{Hagelin}}, \binits{S.}},
\bauthor{\bsnm{{Lascaux}}, \binits{F.}}:
\byear{2010},
\bctitle{{Mt. Graham: Optical Turbulence Vertical Distribution with Standard
  and High Resolution}}.
In: \bbtitle{{Ground-based and Airborne Telescopes III}},
\bsertitle{Proc. SPIE}
\bseriesno{7733},
\bfpage{77331P}.
\doiurl{10.1117/12.856904}.
\end{bchapter}
\endbibitem

\bibitem[\protect\citeauthoryear{{Mathew}, {Zakharov}, and
  {Solanki}}{2009}]{Mathew2009}
\begin{barticle}
\bauthor{\bsnm{{Mathew}}, \binits{S.K.}},
\bauthor{\bsnm{{Zakharov}}, \binits{V.}},
\bauthor{\bsnm{{Solanki}}, \binits{S.K.}}:
\byear{2009},
\batitle{{Stray Light Correction and Contrast Analysis of Hinode Broad-band
  Images}}.
\bjtitle{Astron. Astrophys.}
\bvolume{501},
\bfpage{L19}.
\end{barticle}
\endbibitem

\bibitem[\protect\citeauthoryear{{McInnes}, {Healy}, and
  {Melville}}{2018}]{McInnes2018}
\begin{botherref}
\oauthor{\bsnm{{McInnes}}, \binits{L.}},
\oauthor{\bsnm{{Healy}}, \binits{J.}},
\oauthor{\bsnm{{Melville}}, \binits{J.}}:
2018,
{UMAP: Uniform Manifold Approximation and Projection for Dimension Reduction}.
\textit{arXiv e-prints},
arXiv:1802.03426.
\end{botherref}
\endbibitem

\bibitem[\protect\citeauthoryear{{Moffat}}{1969}]{Moffat1969}
\begin{barticle}
\bauthor{\bsnm{{Moffat}}, \binits{A.F.J.}}:
\byear{1969},
\batitle{{A Theoretical Investigation of Focal Stellar Images in the
  Photographic Emulsion and Application to Photographic Photometry}}.
\bjtitle{Astron. Astrophys.}
\bvolume{3},
\bfpage{455}.
\end{barticle}
\endbibitem

\bibitem[\protect\citeauthoryear{{Nayar} and {Nakagawa}}{1994}]{Nayar1994}
\begin{barticle}
\bauthor{\bsnm{{Nayar}}, \binits{S.Y.}},
\bauthor{\bsnm{{Nakagawa}}, \binits{Y.}}:
\byear{1994},
\batitle{{Shape from Focus}}.
\bjtitle{IEEE Trans. Pattern Anal. Mach. Intell.}
\bvolume{16},
\bfpage{824}.
\doiurl{10.1109/34.308479}.
\end{barticle}
\endbibitem

\bibitem[\protect\citeauthoryear{{Pertuz}, {Puig}, and
  {Garcia}}{2013}]{Pertuz2013}
\begin{barticle}
\bauthor{\bsnm{{Pertuz}}, \binits{S.}},
\bauthor{\bsnm{{Puig}}, \binits{D.}},
\bauthor{\bsnm{{Garcia}}, \binits{M.A.}}:
\byear{2013},
\batitle{{Analysis of Focus Measure Operators for Shape-from-focus}}.
\bjtitle{{Pattern Recognit.}}
\bvolume{46},
\bfpage{1415}.
\doiurl{doi.org/10.1016/j.patcog.2012.11.011}.
\end{barticle}
\endbibitem

\bibitem[\protect\citeauthoryear{{Popowicz}
  \textit{et~al.}}{2017}]{Popowicz2017}
\begin{barticle}
\bauthor{\bsnm{{Popowicz}}, \binits{A.}},
\bauthor{\bsnm{{Radlak}}, \binits{K.}},
\bauthor{\bsnm{{Bernacki}}, \binits{K.}},
\bauthor{\bsnm{{Orlov}}, \binits{V.}}:
\byear{2017},
\batitle{{Review of Image Quality Measures for Solar Imaging}}.
\bjtitle{Solar Phys.}
\bvolume{292},
\bfpage{187}.
\doiurl{10.1007/s11207-017-1211-3}.
\end{barticle}
\endbibitem

\bibitem[\protect\citeauthoryear{{Puschmann} and
  {Sailer}}{2006}]{Puschmann2006a}
\begin{barticle}
\bauthor{\bsnm{{Puschmann}}, \binits{K.G.}},
\bauthor{\bsnm{{Sailer}}, \binits{M.}}:
\byear{2006},
\batitle{{Speckle Reconstruction of Photometric Data Observed with Adaptive
  Optics}}.
\bjtitle{Astron. Astrophys.}
\bvolume{454},
\bfpage{1011}.
\doiurl{10.1051/0004-6361:20053918}.
\end{barticle}
\endbibitem

\bibitem[\protect\citeauthoryear{{Rast}}{1995}]{Rast1995}
\begin{barticle}
\bauthor{\bsnm{{Rast}}, \binits{M.P.}}:
\byear{1995},
\batitle{{On the Nature of 'Exploding' Granules and Granule Fragmentation}}.
\bjtitle{Astrophys. J.}
\bvolume{443},
\bfpage{863}.
\doiurl{10.1086/175576}.
\end{barticle}
\endbibitem

\bibitem[\protect\citeauthoryear{{Ricort} \textit{et~al.}}{1981}]{Ricort1981}
\begin{barticle}
\bauthor{\bsnm{{Ricort}}, \binits{G.}},
\bauthor{\bsnm{{Aime}}, \binits{C.}},
\bauthor{\bsnm{{Roddier}}, \binits{C.}},
\bauthor{\bsnm{{Borgnino}}, \binits{J.}}:
\byear{1981},
\batitle{{Determination of Fried's Parameter r$_{0}$ Prediction for the
  Observed r.m.s. Contrast in Solar Granulation}}.
\bjtitle{Solar Phys.}
\bvolume{69},
\bfpage{223}.
\doiurl{10.1007/BF00149990}.
\end{barticle}
\endbibitem

\bibitem[\protect\citeauthoryear{{Rimmele} and {Marino}}{2011}]{Rimmele2011}
\begin{barticle}
\bauthor{\bsnm{{Rimmele}}, \binits{T.R.}},
\bauthor{\bsnm{{Marino}}, \binits{J.}}:
\byear{2011},
\batitle{{Solar Adaptive Optics}}.
\bjtitle{Liv. Rev. Sol. Phys.}
\bvolume{8},
\bfpage{2}.
\doiurl{10.12942/lrsp-2011-2}.
\end{barticle}
\endbibitem

\bibitem[\protect\citeauthoryear{{Sarazin} and {Roddier}}{1990}]{Sarazin1990}
\begin{barticle}
\bauthor{\bsnm{{Sarazin}}, \binits{M.}},
\bauthor{\bsnm{{Roddier}}, \binits{F.}}:
\byear{1990},
\batitle{{The ESO Differential Image Motion Monitor}}.
\bjtitle{Astron. Astrophys.}
\bvolume{227},
\bfpage{294}.
\end{barticle}
\endbibitem

\bibitem[\protect\citeauthoryear{{Scharmer}
  \textit{et~al.}}{2019}]{Scharmer2019}
\begin{barticle}
\bauthor{\bsnm{{Scharmer}}, \binits{G.B.}},
\bauthor{\bsnm{{L{\"o}fdahl}}, \binits{M.G.}},
\bauthor{\bsnm{{Sliepen}}, \binits{G.}},
\bauthor{\bsnm{{de la Cruz Rodr{\'\i}guez}}, \binits{J.}}:
\byear{2019},
\batitle{{Is the Sky the Limit?. Performance of the Revamped Swedish 1-m Solar
  Telescope and its Blue- and Red-beam Reimaging Systems}}.
\bjtitle{Astron. Astrophys.}
\bvolume{626},
\bfpage{A55}.
\doiurl{10.1051/0004-6361/201935735}.
\end{barticle}
\endbibitem

\bibitem[\protect\citeauthoryear{{Scharr}}{2007}]{Scharr2007}
\begin{bchapter}
\bauthor{\bsnm{{Scharr}}, \binits{H.}}:
\byear{2007},
\bctitle{{Optimal Filters for Extended Optical Flow}}.
In: \beditor{\bsnm{{J{\"a}hne}}, \binits{B.}},
\beditor{\bsnm{{Mester}}, \binits{R.}},
\beditor{\bsnm{{Barth}}, \binits{B.}},
\beditor{\bsnm{{Scharr}}, \binits{H.}} (eds.)
\bbtitle{{Complex Motion}}
\bsertitle{{Lecture Notes in Computer Sciences}}
\bseriesno{3417},
\bpublisher{{Springer}},
\blocation{{Berlin}},
\bfpage{14}.
\doiurl{10.1007/978-540-69866-1{\_}2}.
\end{bchapter}
\endbibitem

\bibitem[\protect\citeauthoryear{{Schmidt} \textit{et~al.}}{2012}]{Schmidt2012}
\begin{barticle}
\bauthor{\bsnm{{Schmidt}}, \binits{W.}},
\bauthor{\bsnm{{von der L{\"u}he}}, \binits{O.}},
\bauthor{\bsnm{{Volkmer}}, \binits{R.}},
\bauthor{\bsnm{{Denker}}, \binits{C.}},
\bauthor{\bsnm{{Solanki}}, \binits{S.K.}},
\bauthor{\bsnm{{Balthasar}}, \binits{H.}},
\bauthor{\bsnm{{Bello Gonzalez}}, \binits{N.}},
\bauthor{\bsnm{{Berkefeld}}, \binits{T.}},
\bauthor{\bsnm{{Collados}}, \binits{M.}},
\bauthor{\bsnm{{Fischer}}, \binits{A.}},
\bauthor{\bsnm{{Halbgewachs}}, \binits{C.}},
\bauthor{\bsnm{{Heidecke}}, \binits{F.}},
\bauthor{\bsnm{{Hofmann}}, \binits{A.}},
\bauthor{\bsnm{{Kneer}}, \binits{F.}},
\bauthor{\bsnm{{Lagg}}, \binits{A.}},
\bauthor{\bsnm{{Nicklas}}, \binits{H.}},
\bauthor{\bsnm{{Popow}}, \binits{E.}},
\bauthor{\bsnm{{Puschmann}}, \binits{K.G.}},
\bauthor{\bsnm{{Schmidt}}, \binits{D.}},
\bauthor{\bsnm{{Sigwarth}}, \binits{M.}},
\bauthor{\bsnm{{Sobotka}}, \binits{M.}},
\bauthor{\bsnm{{Soltau}}, \binits{D.}},
\bauthor{\bsnm{{Staude}}, \binits{J.}},
\bauthor{\bsnm{{Strassmeier}}, \binits{K.G.}},
\bauthor{\bsnm{{Waldmann}}, \binits{T.A.}}:
\byear{2012},
\batitle{{The 1.5 Meter Solar Telescope GREGOR}}.
\bjtitle{Astron. Nachr.}
\bvolume{333},
\bfpage{796}.
\doiurl{10.1002/asna.201211725}.
\end{barticle}
\endbibitem

\bibitem[\protect\citeauthoryear{{Scorer}}{1963}]{Scorer1963}
\begin{bchapter}
\bauthor{\bsnm{{Scorer}}, \binits{R.S.}}:
\byear{1963},
\bctitle{{The Causes of Atmospheric Inhomogeneities}}.
In: \beditor{\bsnm{{R{\"o}sch}}, \binits{J.}} (ed.)
\bbtitle{{Le Choix des Sites d'Observation Astronomiques (Site Testing)}},
\bsertitle{IAU Symp.}
\bseriesno{19},
\bfpage{132}.
\doiurl{10.1017/S0074180900051846}.
\end{bchapter}
\endbibitem

\bibitem[\protect\citeauthoryear{{Street}, {Carroll}, and
  {Ruppert}}{1988}]{Street1988}
\begin{barticle}
\bauthor{\bsnm{{Street}}, \binits{J.O.}},
\bauthor{\bsnm{{Carroll}}, \binits{R.J.}},
\bauthor{\bsnm{{Ruppert}}, \binits{D.}}:
\byear{1988},
\batitle{{A Note on Computing Robust Regression Estimates via Iteratively
  Reweighted Least Squares}}.
\bjtitle{Am. Stat.}
\bvolume{42},
\bfpage{152}.
\doiurl{10.1080/00031305.1988.10475548}.
\end{barticle}
\endbibitem

\bibitem[\protect\citeauthoryear{{Tarbell} and {Smithson}}{1981}]{Tarbell1981}
\begin{bchapter}
\bauthor{\bsnm{{Tarbell}}, \binits{T.}},
\bauthor{\bsnm{{Smithson}}, \binits{R.}}:
\byear{1981},
\bctitle{{A Simple Image Motion Compensation System for Solar Observations}}.
In: \beditor{\bsnm{{Dunn}}, \binits{R.B.}} (ed.)
\bbtitle{{Solar Instrumentation: What's Next?}},
\bconflocation{Sunspot, New Mexico},
\bfpage{491}.
\end{bchapter}
\endbibitem

\bibitem[\protect\citeauthoryear{{Tokovinin}}{2002}]{Tokovinin2002}
\begin{barticle}
\bauthor{\bsnm{{Tokovinin}}, \binits{A.}}:
\byear{2002},
\batitle{{From Differential Image Motion to Seeing}}.
\bjtitle{Publ. Astron. Soc. Pac.}
\bvolume{114},
\bfpage{1156}.
\doiurl{10.1086/342683}.
\end{barticle}
\endbibitem

\bibitem[\protect\citeauthoryear{{Tsuneta} \textit{et~al.}}{2008}]{Tsuneta2008}
\begin{barticle}
\bauthor{\bsnm{{Tsuneta}}, \binits{S.}},
\bauthor{\bsnm{{Ichimoto}}, \binits{K.}},
\bauthor{\bsnm{{Katsukawa}}, \binits{Y.}},
\bauthor{\bsnm{{Nagata}}, \binits{S.}},
\bauthor{\bsnm{{Otsubo}}, \binits{M.}},
\bauthor{\bsnm{{Shimizu}}, \binits{T.}},
\bauthor{\bsnm{{Suematsu}}, \binits{Y.}},
\bauthor{\bsnm{{Nakagiri}}, \binits{M.}},
\bauthor{\bsnm{{Noguchi}}, \binits{M.}},
\bauthor{\bsnm{{Tarbell}}, \binits{T.}},
\bauthor{\bsnm{{Title}}, \binits{A.}},
\bauthor{\bsnm{{Shine}}, \binits{R.}},
\bauthor{\bsnm{{Rosenberg}}, \binits{W.}},
\bauthor{\bsnm{{Hoffmann}}, \binits{C.}},
\bauthor{\bsnm{{Jurcevich}}, \binits{B.}},
\bauthor{\bsnm{{Kushner}}, \binits{G.}},
\bauthor{\bsnm{{Levay}}, \binits{M.}},
\bauthor{\bsnm{{Lites}}, \binits{B.}},
\bauthor{\bsnm{{Elmore}}, \binits{D.}},
\bauthor{\bsnm{{Matsushita}}, \binits{T.}},
\bauthor{\bsnm{{Kawaguchi}}, \binits{N.}},
\bauthor{\bsnm{{Saito}}, \binits{H.}},
\bauthor{\bsnm{{Mikami}}, \binits{I.}},
\bauthor{\bsnm{{Hill}}, \binits{L.D.}},
\bauthor{\bsnm{{Owens}}, \binits{J.K.}}:
\byear{2008},
\batitle{{The Solar Optical Telescope for the Hinode Mission: An Overview}}.
\bjtitle{Solar Phys.}
\bvolume{249},
\bfpage{167}.
\doiurl{10.1007/s11207-008-9174-z}.
\end{barticle}
\endbibitem

\bibitem[\protect\citeauthoryear{{Tyson}}{1998}]{Tyson1998}
\begin{bbook}
\bauthor{\bsnm{{Tyson}}, \binits{R.K.}}:
\byear{1998},
\bbtitle{{Principles of Adaptive Optics}},
\bpublisher{Academic Press},
\blocation{Boston, Massachusetts}.
\end{bbook}
\endbibitem

\bibitem[\protect\citeauthoryear{{van der Maaten} and
  {Hinton}}{2008}]{VanderMaaten2008}
\begin{barticle}
\bauthor{\bsnm{{van der Maaten}}, \binits{L.}},
\bauthor{\bsnm{{Hinton}}, \binits{G.}}:
\byear{2008},
\batitle{{Visualizing Data using t-SNE}}.
\bjtitle{J. Mach. Learn. Res.}
\bvolume{9},
\bfpage{2579}.
\end{barticle}
\endbibitem

\bibitem[\protect\citeauthoryear{{van Noort}, {Rouppe van der Voort}, and
  {L{\"o}fdahl}}{2006}]{vanNoort2006}
\begin{bchapter}
\bauthor{\bsnm{{van Noort}}, \binits{M.}},
\bauthor{\bsnm{{Rouppe van der Voort}}, \binits{L.}},
\bauthor{\bsnm{{L{\"o}fdahl}}, \binits{M.}}:
\byear{2006},
\bctitle{{Solar Image Restoration by use of Multi-object Multi-frame Blind
  Deconvolution}}.
In: \beditor{\bsnm{{Leibacher}}, \binits{J.}},
\beditor{\bsnm{{Stein}}, \binits{R.F.}},
\beditor{\bsnm{{Uitenbroek}}, \binits{H.}} (eds.)
\bbtitle{Solar MHD Theory and Observations: A High Spatial Resolution
  Perspective},
\bsertitle{ASP Conf. Ser.}
\bseriesno{354},
\bfpage{55}.
\end{bchapter}
\endbibitem

\bibitem[\protect\citeauthoryear{{van Noort}
  \textit{et~al.}}{2013}]{vanNoort2013}
\begin{barticle}
\bauthor{\bsnm{{van Noort}}, \binits{M.}},
\bauthor{\bsnm{{Lagg}}, \binits{A.}},
\bauthor{\bsnm{{Tiwari}}, \binits{S.K.}},
\bauthor{\bsnm{{Solanki}}, \binits{S.K.}}:
\byear{2013},
\batitle{{Peripheral Downflows in Sunspot Penumbrae}}.
\bjtitle{Astron. Astrophys.}
\bvolume{557},
\bfpage{A24}.
\doiurl{10.1051/0004-6361/201321073}.
\end{barticle}
\endbibitem

\bibitem[\protect\citeauthoryear{{Verdoni} and {Denker}}{2007}]{Verdoni2007}
\begin{barticle}
\bauthor{\bsnm{{Verdoni}}, \binits{A.P.}},
\bauthor{\bsnm{{Denker}}, \binits{C.}}:
\byear{2007},
\batitle{{The Local Seeing Environment at Big Bear Solar Observatory}}.
\bjtitle{Publ. Astron. Soc. Pac.}
\bvolume{119},
\bfpage{793}.
\end{barticle}
\endbibitem

\bibitem[\protect\citeauthoryear{{Vernin} and {Munoz-Tunon}}{1994}]{Vernin1994}
\begin{barticle}
\bauthor{\bsnm{{Vernin}}, \binits{J.}},
\bauthor{\bsnm{{Munoz-Tunon}}, \binits{C.}}:
\byear{1994},
\batitle{{Optical Seeing at La Palma Observatory. II. Intensive Site Testing
  Campaign at the Nordic Optical Telescope}}.
\bjtitle{Astron. Astrophys.}
\bvolume{284},
\bfpage{311}.
\end{barticle}
\endbibitem

\bibitem[\protect\citeauthoryear{{von der L{\"u}he}}{1984}]{vonderLuehe1984}
\begin{barticle}
\bauthor{\bsnm{{von der L{\"u}he}}, \binits{O.}}:
\byear{1984},
\batitle{{Estimating Fried's Parameter from a Time Series of an Arbitrary
  Resolved Object Imaged through Atmospheric Turbulence}}.
\bjtitle{J. Opt. Soc. Am. A}
\bvolume{1},
\bfpage{510}.
\doiurl{10.1364/JOSAA.1.000510}.
\end{barticle}
\endbibitem

\bibitem[\protect\citeauthoryear{{von der L{\"u}he}}{1993}]{vonderLuehe1993}
\begin{barticle}
\bauthor{\bsnm{{von der L{\"u}he}}, \binits{O.}}:
\byear{1993},
\batitle{{Speckle Imaging of Solar Small Scale Structure. I.\ Methods}}.
\bjtitle{Astron. Astrophys.}
\bvolume{268},
\bfpage{374}.
\end{barticle}
\endbibitem

\bibitem[\protect\citeauthoryear{{von der L{\"u}he}}{1998}]{vonderLuehe1998}
\begin{barticle}
\bauthor{\bsnm{{von der L{\"u}he}}, \binits{O.}}:
\byear{1998},
\batitle{{High-resolution Observations with the German Vacuum Tower Telescope
  on Tenerife}}.
\bjtitle{New Astron. Rev.}
\bvolume{42},
\bfpage{493}.
\doiurl{10.1016/S1387-6473(98)00060-8}.
\end{barticle}
\endbibitem

\bibitem[\protect\citeauthoryear{{von der L{\"u}he} and
  {Dunn}}{1987}]{vonderLuehe1987}
\begin{barticle}
\bauthor{\bsnm{{von der L{\"u}he}}, \binits{O.}},
\bauthor{\bsnm{{Dunn}}, \binits{R.B.}}:
\byear{1987},
\batitle{{Solar Granulation Power Spectra from Speckle Interferometry}}.
\bjtitle{Astron. Astrophys.}
\bvolume{177},
\bfpage{265}.
\end{barticle}
\endbibitem

\bibitem[\protect\citeauthoryear{{von der L{\"u}he}
  \textit{et~al.}}{2003}]{vonderLuehe2003}
\begin{bchapter}
\bauthor{\bsnm{{von der L{\"u}he}}, \binits{O.}},
\bauthor{\bsnm{{Soltau}}, \binits{D.}},
\bauthor{\bsnm{{Berkefeld}}, \binits{T.}},
\bauthor{\bsnm{{Schelenz}}, \binits{T.}}:
\byear{2003},
\bctitle{{KAOS: Adaptive Optics System for the Vacuum Tower Telescope at Teide
  Observatory}}.
In: \beditor{\bsnm{{Keil}}, \binits{S.L.}},
\beditor{\bsnm{{Avakyan}}, \binits{S.V.}} (eds.)
\bbtitle{Innovative Telescopes and Instrumentation for Solar Astrophysics},
\bsertitle{Proc. SPIE}
\bseriesno{4853},
\bfpage{187}.
\doiurl{10.1117/12.498659}.
\end{bchapter}
\endbibitem

\bibitem[\protect\citeauthoryear{{Wang} \textit{et~al.}}{2018}]{Wang2018}
\begin{barticle}
\bauthor{\bsnm{{Wang}}, \binits{Z.}},
\bauthor{\bsnm{{Zhang}}, \binits{L.}},
\bauthor{\bsnm{{Kong}}, \binits{L.}},
\bauthor{\bsnm{{Bao}}, \binits{H.}},
\bauthor{\bsnm{{Guo}}, \binits{Y.}},
\bauthor{\bsnm{{Rao}}, \binits{X.}},
\bauthor{\bsnm{{Zhong}}, \binits{L.}},
\bauthor{\bsnm{{Zhu}}, \binits{L.}},
\bauthor{\bsnm{{Rao}}, \binits{C.}}:
\byear{2018},
\batitle{{A Modified S-DIMM+: Applying Additional Height Grids for
  Characterizing Daytime Seeing Profiles}}.
\bjtitle{Mon. Not. R. Astron. Soc.}
\bvolume{478},
\bfpage{1459}.
\doiurl{10.1093/mnras/sty1097}.
\end{barticle}
\endbibitem

\bibitem[\protect\citeauthoryear{{Wedemeyer-B{\"o}hm}}{2008}]{WedemeyerBoehm2008}
\begin{barticle}
\bauthor{\bsnm{{Wedemeyer-B{\"o}hm}}, \binits{S.}}:
\byear{2008},
\batitle{{Point Spread Functions for the Solar Optical Telescope onboard
  Hinode}}.
\bjtitle{Astron. Astrophys.}
\bvolume{487},
\bfpage{399}.
\doiurl{10.1051/0004-6361:200809819}.
\end{barticle}
\endbibitem

\bibitem[\protect\citeauthoryear{{Wedemeyer-B{\"o}hm} and {Rouppe van der
  Voort}}{2009}]{WedemeyerBoehm2009b}
\begin{barticle}
\bauthor{\bsnm{{Wedemeyer-B{\"o}hm}}, \binits{S.}},
\bauthor{\bsnm{{Rouppe van der Voort}}, \binits{L.}}:
\byear{2009},
\batitle{{On the Continuum Intensity Distribution of the Solar Photosphere}}.
\bjtitle{Astron. Astrophys.}
\bvolume{503},
\bfpage{225}.
\doiurl{10.1051/0004-6361/200911983}.
\end{barticle}
\endbibitem

\bibitem[\protect\citeauthoryear{{Weigelt} and {Wirnitzer}}{1983}]{Weigelt1983}
\begin{barticle}
\bauthor{\bsnm{{Weigelt}}, \binits{G.}},
\bauthor{\bsnm{{Wirnitzer}}, \binits{B.}}:
\byear{1983},
\batitle{{Image Reconstruction by the Speckle-masking Method}}.
\bjtitle{Opt. Lett.}
\bvolume{8},
\bfpage{389}.
\end{barticle}
\endbibitem

\bibitem[\protect\citeauthoryear{{Xue} \textit{et~al.}}{2014}]{Xue2014}
\begin{barticle}
\bauthor{\bsnm{{Xue}}, \binits{W.}},
\bauthor{\bsnm{{Zhang}}, \binits{L.}},
\bauthor{\bsnm{{Mou}}, \binits{X.}},
\bauthor{\bsnm{{Bovik}}, \binits{A.C.}}:
\byear{2014},
\batitle{{Gradient Magnitude Similarity Deviation: A Highly Efficient
  Perceptual Image Quality Index}}.
\bjtitle{IEEE Trans. Image Process.}
\bvolume{23},
\bfpage{684}.
\doiurl{10.1109/TIP.2013.2293423}.
\end{barticle}
\endbibitem

\bibitem[\protect\citeauthoryear{{Zirin} and {Mosher}}{1988}]{Zirin1988b}
\begin{barticle}
\bauthor{\bsnm{{Zirin}}, \binits{H.}},
\bauthor{\bsnm{{Mosher}}, \binits{J.M.}}:
\byear{1988},
\batitle{{The Caltech Solar Site Survey, 1965--1967}}.
\bjtitle{Solar Phys.}
\bvolume{115},
\bfpage{183}.
\doiurl{10.1007/BF00146239}.
\end{barticle}
\endbibitem

\end{thebibliography}
\end{document}